\documentclass[numberedappendix,useAMS,usenatbib,letterpaper]{mn2e}
\usepackage[pass]{geometry}
\usepackage{amssymb,amsmath}
\usepackage{verbatim}
\usepackage{color}
\usepackage{hyperref}
\usepackage{booktabs}
\usepackage{float}
\usepackage{needspace}
\usepackage{enumitem}
\usepackage{subfig}
\usepackage{graphicx}
\definecolor{linkcolor}{rgb}{0,0,0.25}
\hypersetup{
  colorlinks=true,        
  linkcolor=linkcolor,    
  citecolor=linkcolor,    
  filecolor=linkcolor,    
  urlcolor=linkcolor      
}
\newcounter{address}
\DeclareMathAlphabet{\mathsc}{OT1}{cmr}{m}{sc}
\def\testbx{bx}%
\DeclareRobustCommand{\ion}[2]{%
\relax\ifmmode
\ifx\testbx\f@series
{\mathbf{#1\,\mathsc{#2}}}\else
{\mathrm{#1\,\mathsc{#2}}}\fi
\else\textup{#1\,{\mdseries\textsc{#2}}}%
\fi}

\title[Billion particle mergers]
{Resolving local and global kinematic signatures of satellite mergers with billion particle simulations}

\author[J. A. S. Hunt et al.]
{\parbox{\textwidth}{Jason A.~S.~Hunt$^1$, Ioana A. Stelea$^2$, Kathryn V. Johnston$^{2,1}$, Suroor S. Gandhi$^{3}$, Chervin F. P. Laporte$^{4,5}$ \& Jeroen B{\'e}dorf\hspace{0.08cm}$^{6}$}%
\\
\\
$^{1}$ Center for Computational Astrophysics, Flatiron Institute, 162 5th Av., New York City, NY 10010, USA\\
$^2$ Department of Astronomy, Columbia University, New York, NY 10027, USA\\
$^3$ Center for Cosmology and Particle Physics, Department of Physics, New York University, 726 Broadway, New York, NY 10003, USA\\
$^{4}$ Institut de Ci\`encies del Cosmos (ICCUB), Universitat de Barcelona (IEEC-UB), Mart\'i i Franqu\`es 1, 08028 Barcelona, Spain\\
$^{5}$ Kavli Institute for the Physics and Mathematics of the Universe (WPI), The University of Tokyo Institutes for Advanced Study (UTIAS),\\ \hspace{0.22cm}The University of Tokyo, Chiba 277-8583, Japan \\
$^6$ Leiden Observatory, Leiden University, P.O. Box 9513, 2300 RA Leiden, the Netherlands\\
}
\pagerange{\pageref{firstpage}--\pageref{lastpage}}
\pubyear{2020}

\begin{document}

\maketitle

\label{firstpage}

\begin{abstract}
In this work we present two new $\sim10^9$ particle self-consistent simulations of the merger of a Sagittarius-like dwarf galaxy with a Milky Way-like disc galaxy. One model is a violent merger creating a thick disc, and a Gaia-Enceladus/Sausage like remnant. The other is a highly stable disc which we use to illustrate how the improved phase space resolution allows us to better examine the formation and evolution of structures that have been observed in small, local volumes in the Milky Way, such as the $z-v_z$ phase spiral and clustering in the $v_{\mathrm{R}}-v_{\phi}$ plane when compared to previous works. The local $z-v_z$ phase spirals are clearly linked to the global asymmetry across the disc: we find both 2-armed and 1-armed phase spirals, which are related to breathing and bending behaviors respectively. Hercules-like moving groups are common, clustered in $v_{\mathrm{R}}-v_{\phi}$ in local data samples in the simulation. These groups migrate outwards from the inner galaxy, matching observed metallicity trends even in the absence of a galactic bar. We currently release the best fitting `present day' merger snapshots along with the unperturbed galaxies for comparison.
\end{abstract}


\section{Introduction}\label{intro}
The Milky Way has long been known to be a barred spiral galaxy \cite[e.g.][]{BS91}, but our location within it leads to complex observational selection effects such as those imposed by the dust extinction, which make it difficult to construct a global picture of our home galaxy. Thus, the observational data has long been complemented by the construction of various Milky Way models, to help us understand the observational biases and reveal the Galaxy beyond. However, such modelling frequently assumes that the system is in equilibrium, which is a necessary (or significantly simplifying) assumption for many modelling techniques.

The second data release \citep[DR2;][]{DR2} from the European Space Agency's $Gaia$ mission \citep{GaiaMission} highlighted just how far out of equilibrium our Galaxy is. For example, \cite{Antoja+18} showed for the first time the striking spiral, or snail shaped pattern in the $z-v_z$ plane, henceforth phase spiral. This phase spiral has been proposed to arise from the interaction of the Milky Way with the Sagittarius dwarf galaxy, henceforth Sgr \citep[e.g.][]{Antoja+18,LMJG19,Khanna+19}. However, \cite{BB21} find that the amplitude and wavelength of the vertical asymmetry in the Solar neighbourhood cannot be produced by Sgr alone, for a variety of Milky Way potentials, and Sgr orbits and masses. One alternative is that the phase spiral can be created by the buckling of the Galactic bar \citep{Khoperskov+19a}, although the lack of an age dependence in the phase spiral argues against this origin \citep{LMJG19}. 

\begin{table*}
\caption{Halo tidal radius $\epsilon_{\mathrm{h}}$, halo characteristic velocity $\sigma_{\mathrm{h}}$, halo scale length $a_{\mathrm{h}}$, disc mass $M_{\mathrm{d}}$, disc scale length $R_{\mathrm{d}}$, disc scale height $h_{\mathrm{d}}$, bulge tidal radius $\epsilon_{\mathrm{b}}$, characteristic bulge velocity $\sigma_{\mathrm{b}}$, bulge scale length $a_{\mathrm{b}}$ and Toomre parameter $Q$ for the Milky Way like disc galaxy initial conditions for the \texttt{galactics} initial condition generator, as taken from \citet{WD05}. The initial conditions assume $G=1$, with units of kpc, 100 km s$^{-1}$, and $2.33\times10^9$ $M_{\odot}$.}
\begin{tabular}{@{}lllllllllll@{}}
\toprule
Model & $\epsilon_{\mathrm{h}}$ & $\sigma_{\mathrm{h}}$ & $a_{\mathrm{h}}$ & $M_{\mathrm{d}}$ & $R_{\mathrm{d}}$ & $h_{\mathrm{d}}$  & $\epsilon_{\mathrm{b}}$ & $\sigma_{\mathrm{b}}$ & $a_{\mathrm{b}}$ & Q \\ 

\midrule 
M1, D1 & 0.11 & 3.447 & 8.818 & 14.47 & 2.817 & 0.439 & 0.209 & 4.357 & 0.884 & 2.2  \\
M2, D2 & 0.17 & 2.496 & 12.96 & 19.66 & 2.806 & 0.409 & 0.213 & 4.444 & 0.788 & 1.3  \\
\bottomrule
\end{tabular}
\label{params}
\end{table*}

$Gaia$ DR2 also revealed ridge like structures in the $R-v_{\phi}$ plane \citep{Antoja+18,KBCCGHS18,RAF18}, an extension of the long known structure in local velocity space across several kpc. The ridges in the $R-v_{\phi}$ plane, or the local $v_{\mathrm{R}}-v_{\phi}$ plane can be explained by bar resonances \citep[e.g.][]{Fragkoudi+19}, spiral structure \citep[e.g.][]{Hunt+18} or the influence of an external perturber such as Sgr \citep[e.g.][]{LMJG19,Khanna+19}. The true cause is likely a complex combination of the three, which is non-trivial to disentangle.

To model the phase spiral, \cite{BS18} employ a simple impulse approximation model of a satellite passing vertically through the galactic disc and reproduce a qualitatively similar phase spiral to that seen in the $Gaia$ data, for an impact $400\pm150$ Myr ago, which could correspond to the previous pericentric passage of Sgr \citep{DL17}. However, as they are careful to note, this neglects both the geometry of the orbit and perturbation of the frequencies by the impactor. \cite{BS18} also note that to reach the same resolution of the phase spirals in a particle based simulation $\geq2\times10^8$ disc particles would be required, which would be computationally expensive in a fully self-consistent simulation, especially when also modelling the dark matter halo. 

\begin{figure*}
    \centering
    \includegraphics[width=\hsize]{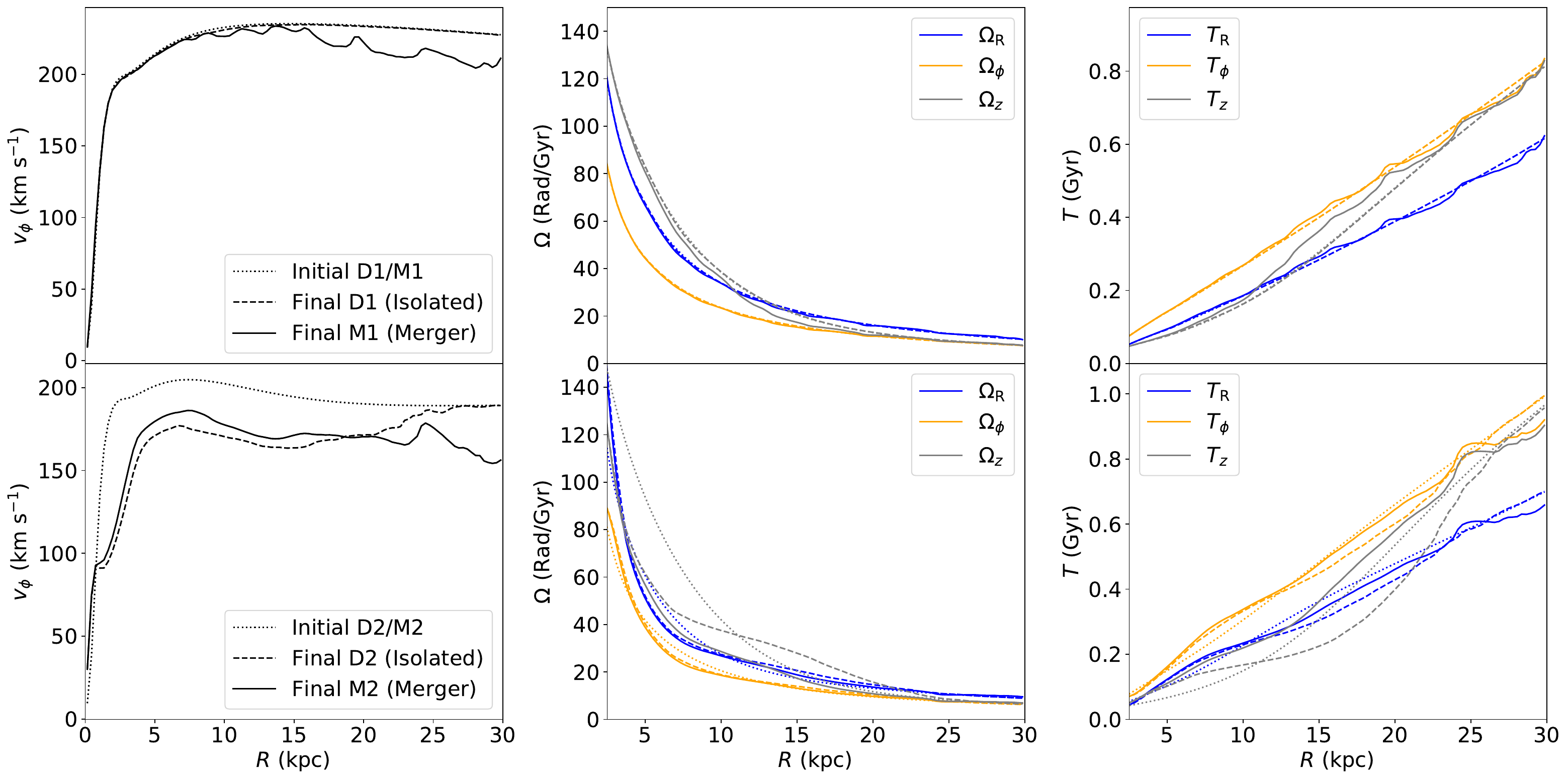}
    \caption{Rotation curve (left column), orbital frequencies (centre column) and orbital period (right column) for Model D1 \& M1 (upper row) and model D2 \& M2 (lower row). The dotted lines show the initial condition, the dashed lines show the `present day' snapshots of the isolated discs D1 \& D2, and the solid lines show the `present day' snapshots of Models M1 \& M2. In the centre and right hand columns, the radial, azimuthal and vertical frequencies and periods are blue, orange and grey respectively.}
    \label{freqs}
\end{figure*}

\begin{figure}[h!]
    \centering
    \includegraphics[width=\hsize]{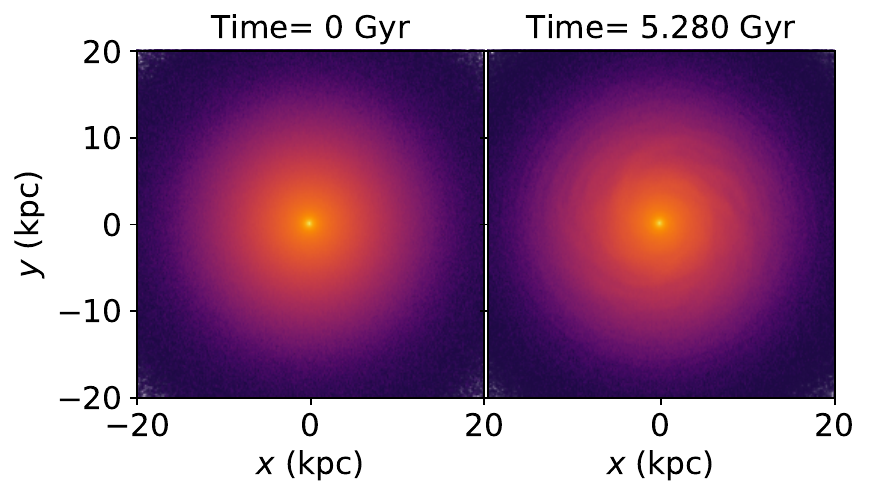}
    \includegraphics[width=\hsize]{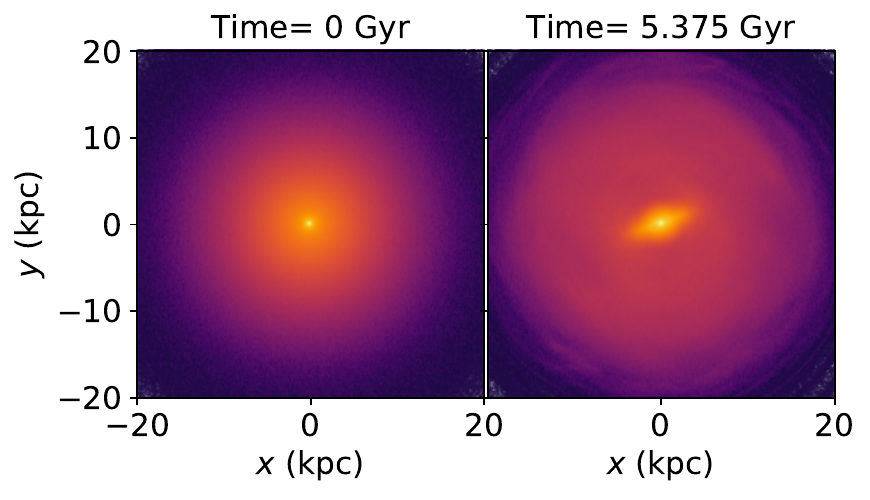}
    \caption{Face on view of the isolated disc Models D1 (upper row) and D2 (lower row) at $t=0$ (left column) and at the end of the simulation (right column).}
    \label{IsoMorph}
\end{figure}

Such fully self-consistent merger simulations \citep[e.g.][]{LJGG-CB18} better capture the real physics of such an interaction, with a dynamically consistent orbit, including dynamical friction and satellite mass loss, along with an extended density profile for the satellite. Such models can reproduce large scale disc dynamics and local kinematic features similar to those observed in the Milky Way \citep[e.g.][]{LMJG19}, but lack the particle resolution to distinguish fine detail over a local volume at a level comparable to current data sets. \cite{BH-TG+21} make a compromise by subjecting a live $7.45\times10^7$ particle galaxy to a point mass perturber, and find a complex response in the disc, consisting of the superposition of a density wave and a corrugated bending wave, which wrap at different rates, giving rise to phase spirals that last for multiple Gyr. However, even the $5\times10^7$ particles in their disc leaves the simulation on the edge of resolving fine features in the outer disc.

In particular, this is the part of the galaxy which will be heavily affected by interaction with a satellite \citep[e.g.][]{Kazantzidis09, Gomez17} which is also becoming increasingly accessible to observed data sets \citep[e.g.][]{AntojaeDR321, L21}. Due to the longer orbital timescales in the outer disc, older excited structures from prior interactions can remain coherent to the present-day. For example, \cite{laporte20} confirmed the Anticenter stream \citep{Grillmair06} to be a fossil relic of a past satellite interaction due to its large composition of predominantly old low-$\alpha$ stars when compared to the bulk of the disc at similar distances. Such mergers have important implications for the growth rate of the Galactic disc and its chemical evolution, making relics from past satellite interactions sensitive probes of the formation history of the Galaxy.

Currently, one of the limiting factors in our interpretation of simulations on local scales is the particle resolution. While the global signatures in such simulations can be easily resolved \citep[e.g.][]{LJGG-CB18,BH-TG+21}, performing analyses of projections such as the $v_{\mathrm{R}}-v_{\phi}$ plane or the $z-v_z$ plane is difficult without increasing the volume of the `local' sample, which in turn dilutes the signal. With $Gaia$ DR2 we have entered a new regime where we have significantly more stars within a few hundred parsecs, with excellent astrometry and radial velocities, than are present in the same volume in the current generation of simulations.

While it is not currently feasible to construct models with the same number of particles as there are stars in our Galaxy, if we desire to improve local phase space resolution we have some options which enable us to get closer. Firstly we can use test particle simulations which require a fraction of the computational cost and enable us to precisely set the galactic parameters, but lack realistic gravitational dynamics. Secondly, we can use GPU accelerated $N$-body codes, such as \texttt{Bonsai} \citep{Bonsai}, which has achieved a 242 billion particle Milky Way model while continuing to scale well \citep{Bonsai-242bil}. Other options would include `Hybrid' simulations which add test particles to a self-consistent model \citep[e.g.][]{QDBMC11} to increase phase space resolution, but not the force resolution, or the use of the Basis Function Expansion (BFE) technique \citep[e.g.][]{CB72,W99} where the force calculation time scales with $\mathcal{O}(N)$, instead of the $\mathcal{O}(N\log N)$ of tree codes, and retains comparable or superior force resolution \citep{PWK21}. More realistic models which include gas, star formation and feedback further increase the computational cost, and are thus by necessity lower resolution than can be attained by pure $N$-body simulations regardless of the simulation technique. However, recent cosmological zoom simulations are achieving impressive resolutions, e.g. \cite{Grand+21} re-simulate a galaxy from the \texttt{Auriga} cosmological simulation \citep{Grand-Auriga} with $\sim10^8$ stellar particles in the main Milky Way-like galaxy.

In this work we concern ourselves less with matching the Milky Way - Sgr system, and more on resolving the response of a self-gravitating disc to a satellite perturber. Thus, we present high resolution ($>1\times10^9$ particle) self-consistent simulations of two mergers, each between a Milky Way like disc galaxy and a dwarf, evolved with \texttt{Bonsai}, which pass the $2\times10^8$ disc particle threshold suggested by \cite{BS18}. We use them to illustrate local dynamical signatures that would be missed in lower resolution models. They are available online at Flathub\footnote{\url{https://flathub.flatironinstitute.org/jhunt2021}}, and this paper serves as the reference document for the simulations, along with a brief analysis of the disc dynamics. In Section \ref{sim} we describe the setup of the simulations. In Section \ref{M1} we examine the structure and dynamics of the first galaxy, focusing on the $z-{v_z}$ phase spiral in Sections \ref{localspiral} and \ref{globalspiral}, and the planar dynamics in Section \ref{UV}. In Section \ref{M2} we examine the second merger simulation. In Section \ref{conclusions} we give our conclusions. 

\section{The simulations}\label{sim}
In this section we describe the setup and evolution of the Milky Way-like disc galaxies which we use as the host galaxies for the merger simulations, the satellite used for the merger, and their combination.

\subsection{The isolated host galaxy simulations, Models D1 and D2}
The initial conditions for the Milky Way like host galaxies, models D1 and D2 were created with the parallelised version of the \texttt{galactics} initial condition generator\footnote{\url{https://github.com/treecode/galactics.parallel}} \citep{KD95}. We chose to use the parameters from the MWa and MWb Milky Way like models from \cite{WD05}, which are summarized in Table \ref{params} as they are already well explored and the purpose of the paper is to examine the merger driven structure and kinematics, not the construction of new best fitting Milky Way like disc galaxies. 

Model D1 \citep[MWb in][]{WD05} is a disc with a high Toomre parameter $Q=2.3$ \citep{Toomre64} which is stable against bar and spiral formation for several Gyr. While such a high stability disc is not representative of the Milky Way, we consider it ideal when constructing a model in which the only non-axisymmetric structure and kinematics will form from interactions with the perturbing dwarf. For comparison, Model D2 \citep[MWa in][]{WD05} has a heavier disc and $Q=1.3$, leaving it stable, but not unreasonably so \citep[see][for a detailed analysis of the models as isolated discs]{WD05}. 

We use the GPU based $N$-body tree code \texttt{Bonsai} \citep{Bonsai,Bonsai-242bil} to evolve Models D1 \& D2 for $\sim5$ Gyr, using a smoothing length of 50 pc and an opening angle $\theta_{\mathrm{o}}=0.4$ radians. Table \ref{ncomps} shows the particle numbers and mass resolution per component for the models. While ideally we would like to have the same mass resolution in each component to avoid heating concerns, we are memory limited to around $1.3\times10^9$ particles on our current system, and we prioritise having at least $2\times10^{8}$ particles in the disc to increase local phase space resolution.

While we do not wish to perform a detailed re-analysis of existing models from the literature, Figure \ref{freqs} shows the rotation curve (left column), along with the frequencies (middle column) and periods (right column) in the radial (blue), vertical (grey) and azimuthal (orange) directions, for the initial discs (dotted lines) and the final state of the isolated discs (dashed lines) for Models D1 (top row) and D2 (bottom row).

The top row shows that Model D1, the high stability disc, experiences almost no change in rotation curve, frequency or period over the course of the $\sim$5 Gyr of isolated evolution, in such that the dashed lines consistently almost cover the dotted lines. Similarly, the top row of Figure \ref{IsoMorph} shows that Model D1 remains a smooth disc for several Gyr of evolution, with some very light spiral structure by the final snapshot at $t=5.28$ Gyr. As such, there is no defined `present day' snapshot as the disc is never a good representation of the morphology of the Milky Way. It is included in the paper and hosted on Flathub as Model D1, snapshot 200, purely to facilitate comparison with the M1 merger model presented later for which it is used as the host galaxy. We use the snapshot at $\sim2$ Gyr, which is ample time for the minor instabilities to have mixed away, and before the emergence of the mild spiral structure later on.

However, the lower row of Figure \ref{freqs} shows that model D2, experiences significant evolution in its rotation curve, orbital frequencies and periods as an isolated galaxy. Morphologically, the lower row of Figure \ref{IsoMorph} shows that the initial condition develops into a barred galaxy with some weak spiral structure. We rotate the bar angle to approximately match the estimates of the Milky Way bar angle with respect to the Solar position, assumed here to be $(x_{\odot}, y_{\odot})=(-8.178, 0)$ kpc. At the final step at $t=5.375$ Gyr, the pattern speed is $\Omega_{\mathrm{b}}=26.7$ km s$^{-1}$ kpc$^{-1}$, which is significantly slower than current estimates of the Milky Way's bar pattern speed which appear to be converging around 37-42 km s$^{-1}$ kpc$^{-1}$ \citep[e.g.][]{BLHM+19,SSE19,Clarke+19}. Thus, the bar does not well match our current view of the Milky Way bar, but again, that is not the purpose of this work, and we host the final snapshot\footnote{Model D2, snapshot 549 on Flathub} only for comparison with the merger model described below.

\begin{figure}
    \centering
    \includegraphics[width=\hsize]{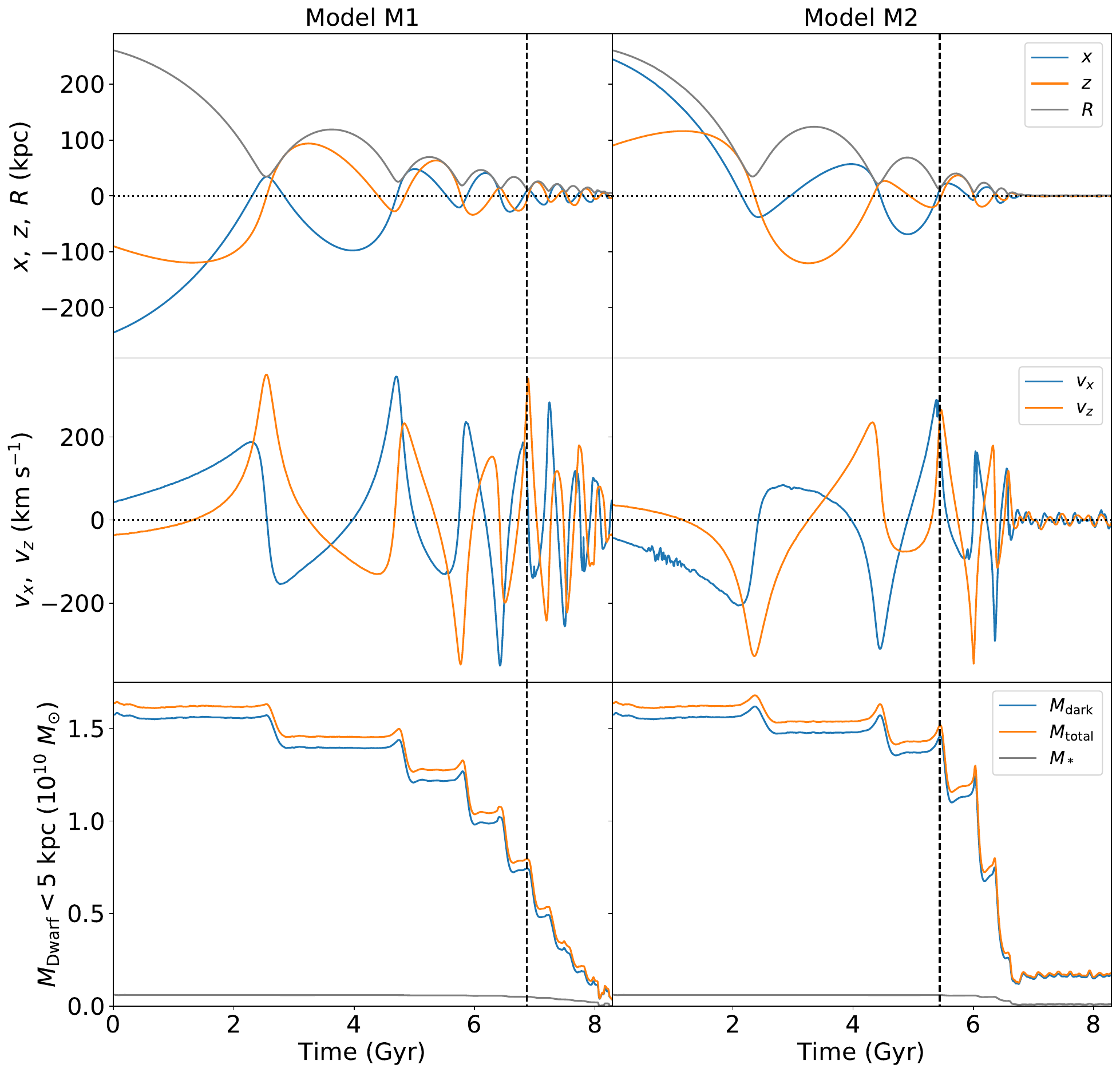}
    \caption{Orbital trajectory in spherical radius $R$ (grey), $x$ (blue) and $z$ (orange, upper row) and velocity $v_x$ (blue) and $v_z$ (orange, middle row), and mass within 5 kpc for the dwarf in $10^{10}$ $M_{\odot}$, for stellar (grey), dark (blue) and total mass (orange, lower row) for Model M1 (left column) and M2 (right column). Note that $y$ and $v_y$ are not shown as the orbit is almost radial in the $x-z$ plane. In model M2 the satellite is fully disrupted around 6.75 Gyr. The vertical line shows the time chosen for the present day.}
    \label{orbits}
\end{figure}

\begin{table}
\caption{Particle number and mass resolution of the components of the various models described in Section \ref{sim}.}
\begin{tabular}{@{}llll@{}}
\toprule
Model & Component & $N_{\mathrm{p}}$ & $M_{\mathrm{p}}\ (M_{\odot})$ \\ \midrule
\midrule
M1, D1 & Disk & 219,607,640 & 160 \\
M1, D1 & Bulge & 21,960,680 & 540 \\ 
M1, D1 & Halo & 878,431,120 & 640 \\ 
M1 & Dwarf stellar & 153,599 & 4,236 \\
M1 & Dwarf dark & 2,880,684 & 19,960 \\ 
M1 & Total & 1,123,033,723 & - \\
M1, D1 & Total host & 1,119,999,440 & - \\
\midrule
M2, D2 & Disk & 250,980,160 & 185 \\
M2, D2 & Bulge & 25,097,920 & 482 \\ 
M2, D2 & Halo & 1,003,921,280 & 674 \\ 
M2 & Dwarf stellar & 153,599 & 4,236 \\
M2 & Dwarf dark & 2,880,684 & 19,960 \\ 
M2 & Total & 1,283,187,242 & - \\
M2, D2 & Total Host & 1,280,152,959 & - \\
\bottomrule
\end{tabular}
\label{ncomps}
\end{table}

\subsection{Dwarf galaxy properties}\label{dwarf}
For simplicity, we use the initial condition for the Sagittarius-like dwarf galaxy from model L2 of \cite{LJGG-CB18}, henceforth called `the dwarf', which is comprised of two Hernquist spheres \citep{H90}. The first represents the dark matter, and has virial mass $M_{200}=6\times10^{10}\ M_{\odot}$, concentration parameter $c_{200}=28$, halo mass, $M_h=8\times10^{10}\ M_{\odot}$ and scale radius $a_h=8$ kpc. The second represents the stellar component embedded in the dark halo, where the stellar mass $M_*=6.4\times10^8\ M_{\odot}$ and $a_h=0.85$ kpc \citep[see][for a more thorough description]{LJGG-CB18}.

\begin{table*}
\caption{Sagittarius position and motion from the literature, and the location of the dwarfs at the `Present day' model snapshots.}
\begin{tabular}{@{}lllllllll@{}}
\toprule
Model/Catalogue & $x$ (kpc) & $y$ (kpc) & $z$ (kpc) & $v_x$ (km s$^{-1}$) & $v_y$ (km s$^{-1}$) & $v_z$ (km s$^{-1}$) \\ \midrule
\citet{Vasiliev+20} & 17.5 & -2.5 & -6.5 & 237.9 & 24.3 & 209.0 \\
M1 & 9.5 & -0.4 & -6.7 & 115.7 & 5.9 & 311.9 \\
M2 & 12.9 & -0.5 & -5.0 & 208.6 & 7.3 & 255.4 \\ 
\bottomrule
\end{tabular}
\label{sgrpos}
\end{table*}

\subsection{Merger simulations, Models M1 and M2}\label{merger}
Model M1 is the merger of the dwarf into the D1 Milky Way model. Model M2 is the merger of the dwarf into the D2 Milky Way model. 

The top left panel of Figure \ref{orbits} shows the orbit of the dwarf in $R$, $x$ and $z$ in Model M1 (referring to the combined system of D1 and the dwarf), with the horizontal dotted line marking the $R, x, z=0$ line. The centre left panel of Figure \ref{orbits} shows $v_x$ and $v_z$ for the dwarf. Note that $y$ and $v_y$ are not shown as the merger is almost radial along the $y$ axis. The lower left panel shows the stellar (grey), dark (blue) and total (orange) mass within 5 kpc of the centre of the dwarf over the course of the simulation. The dwarf loses mass over time, with the `step' like structure corresponding to the pericentric passages, as expected. 

The right hand column of Figure \ref{orbits} shows the same but for Model M2. The dwarf in model M2 initially loses mass at a slower rate than M1, but becomes fully disrupted around 6.75 Gyr following rapid mass loss in the fourth and fifth pericentric passages (note that the mass does not go to zero at $t>6.75$ as the mass within 5 kpc of the median position of dwarf stars is still substantial in the inner galaxy; this is a simple spatial selection to illustrate mass loss, not a rigorous census of bound stars).

We assess the orbital trajectories of the dwarf in Model M1 and M2, and define the `present day' snapshots as the times when the dwarf is closest to the location and motion of Sgr in the Milky Way as given by \cite{Vasiliev+20}. For Model M1 we identify the most `present day' like snapshot as occurring at $t=6.87$ Gyr, and for Model M2 we identify the most `present day' like snapshot as occurring at $t=5.44$ Gyr. The `present day' for both models is marked on Figure \ref{orbits} with a vertical dashed line.

Table \ref{sgrpos} gives the position and motion of Sgr \citep{Vasiliev+20} and the dwarf in Models M1 \& M2. For both models, as in the real MW-Sgr system, the dwarf is approaching a disc crossing, approximately 6 kpc beneath the disc plane, and on the far side of the galaxy from the Sun (considered to be at $(x_{\odot},y_{\odot})=(-8.178, 0)$ kpc), moving away from us roughly along the $x$ axis. However, in both models, our present day snapshots have the dwarf too close to the galactic centre ($R=11.6$ and 13.8 kpc, compared to $R_{\mathrm{Sgr}}=18.8$ kpc). This is because at no point during the simulation are the dwarfs at the correct radius $and$ the correct orbital phase. E.g. we prioritised the phase of the orbit over the distance from the galactic centre.

In addition, at these `present day' snapshots the dwarf is significantly more massive than the remnant mass of $M_{\mathrm{Sgr}}\sim4\times10^8$ $M_{\odot}$ inferred from the central velocity distribution \citep{Vasiliev+20}, and in each case the dwarf is significantly within the inner galaxy before the mass is comparable to that of the Sgr remnant. We are not concerned by the discrepancy as the purpose of these models is to investigate the response of the discs to the satellite perturber, and while such a response may be stronger in the simulation than the true Milky Way, the dynamics should be qualitatively comparable. As such, we defer our attempt to better fit the Milky Way - Sgr system to future work \citep{BBH21}.

For the calculation of the orbits, and the subsequent analysis of the dynamics in later sections we re-center the simulations on the centre of the bulge at each time step as the host galaxies experience reflex motion owing to the in-falling satellites. We also realign the $z$-axis with the galaxy's axis of rotation, which can change over the course of the interaction, especially in the more violent merger of Model M2.

In summary, Model M1 is designed to be a laboratory for exploration of the formation and evolution of the $z-v_z$ phase spiral in an otherwise smooth disc. Model M2 is an example of a violent merger between a Milky Way disc, and a large compact satellite. Future work will include construction of a closer representation of the Milky Way - Sgr system \citep{BBH21} but that is not the focus of this work, which investigates generic merger signatures.

\begin{figure*}
    \centering
    \includegraphics[width=\hsize]{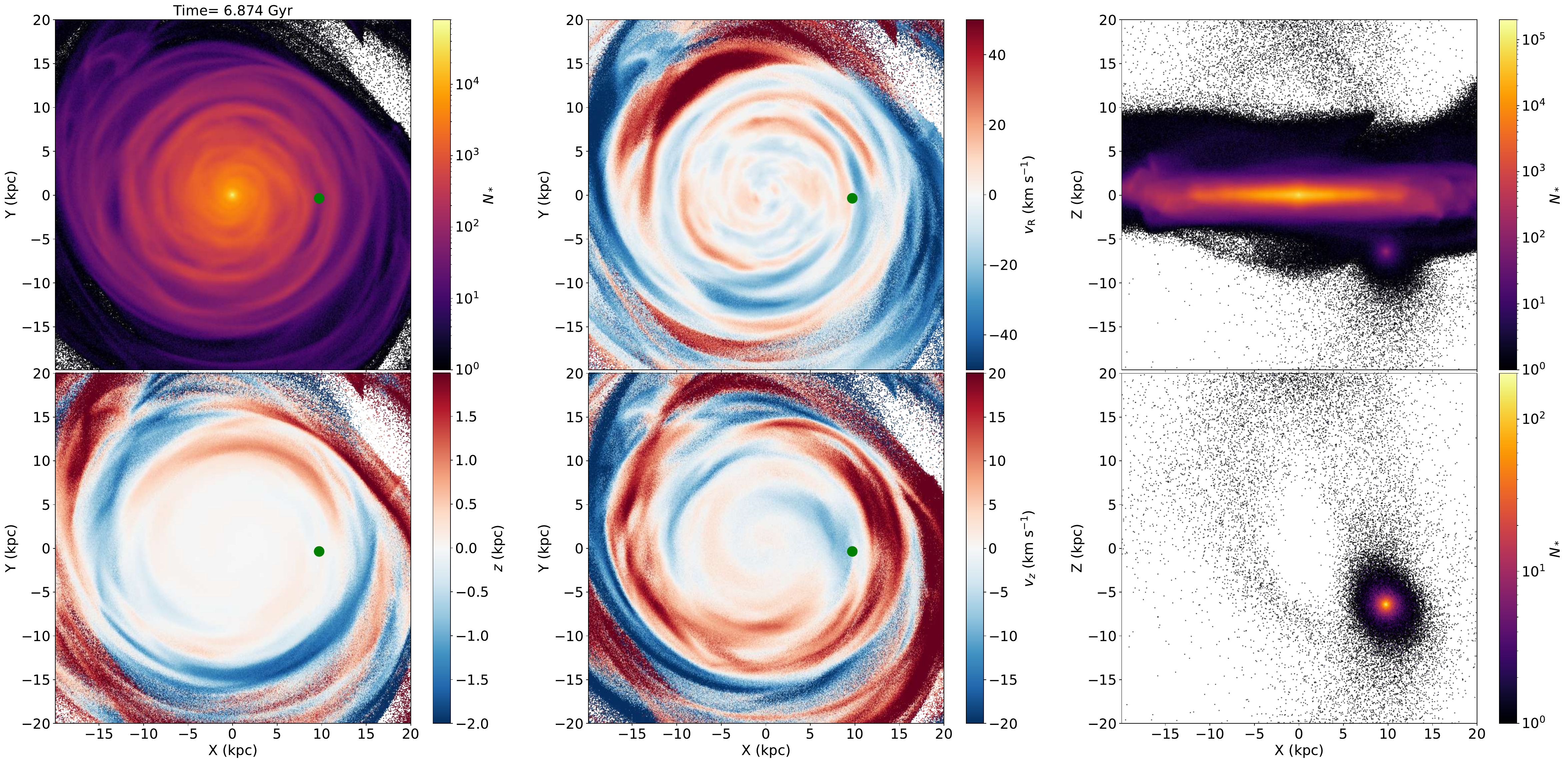}
    \caption{Face on (left and centre columns) and edge on (right column) view of the disc colored by number count (upper left panel and right column), mean radial velocity (upper centre), mean vertical position (lower left) and mean vertical velocity (lower centre) for Model M1 at $t=6.874$ Gyr. The dwarf particles are only included in the right hand column, but its position is marked with a green dot in the face on panels.}
    \label{M1-global}
\end{figure*}

\section{Model M1 structure and kinematics}\label{M1}

Model M1 is designed to be a laboratory for exploration of the formation and evolution of disk substructure in an otherwise smooth disc.

\subsection{Global snapshot of the `present day' state}

Figure \ref{M1-global} shows the stellar morphology and kinematics of Model M1 at the `present day' of $t=6.87$ Gyr\footnote{Model M1, snapshot 703 on Flathub}. The top left panel shows the face on view of the surface density of the disc and dwarf, the top middle shows the mean radial velocity, the top right panel shows surface density in the edge on view of the disc and dwarf, the lower right panel shows the morphology of the stellar component of the dwarf only. The lower left panel shows the mean vertical position, and the lower centre panel shows the mean vertical velocity. The dwarf particles are only shown in the edge on views in the right hand column, although its position is marked in the face on views with a green dot. The downwards pull of the current passage of the dwarf is visible in the vertical velocity signature in the lower centre panel \citep[see][for a further discussion of this effect]{Gandhi+20}.

As one can see from the top left panel the `present day' snapshot of model M1 contains no bar. It does develop one later in the simulation, and a slight hint of the emerging quadrupole can be seen in the inner region of the top centre panel, but it only becomes substantial once the dwarf is in the inner region of the galaxy (see here\footnote{\url{https://users.flatironinstitute.org/~jhunt/M1-Morphology+Kinematics.mp4} (or .gif if mp4 is unsupported)} for the full evolution of Model M1). Model M1 does contain numerous thin, low pitch angle spiral arms which are tidally induced by the interaction with the dwarf. Thus, model M1 is an excellent laboratory to study the structural and kinematic features which arise from a satellite perturber, without contamination from bar resonances. However, it is not designed to be a `best fit' model of the current state of the Milky Way, and the lack of bar, high $Q$, and the location of the dwarf compared to the galactic centre mean that it should not be taken as such. 

The solid lines in the top row of Figure \ref{freqs} show the rotation curve, orbital frequencies and periods for the `present day' snapshot. The rotation has fallen slightly in the outer galaxy, with the bumps corresponding to kinematic substructure induced by the satellite. The radial and azimuthal frequencies and periods do not change much in the highly stable disc, but the vertical frequencies (periods) drop (rise) as the disc is heated by the merging satellite.

\subsection{Model M1: The phase spiral locally}\label{localspiral}
As first highlighted by \cite{Antoja+18} with $Gaia$ DR2, the vertical perturbation of the Galactic disc manifests as a spiral or snail shape in the $z-v_z$ plane. Dynamically, this can be attributed to the influence of an external perturber, or the buckling of the Galactic bar. 

\begin{figure}
    \centering
    \includegraphics[width=\hsize]{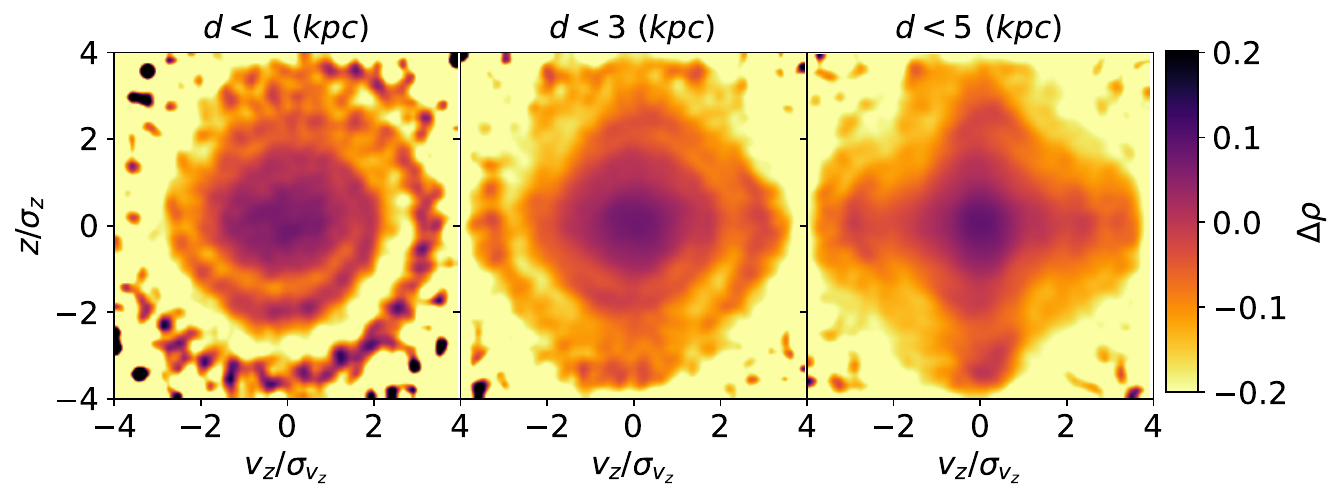}
    \includegraphics[width=\hsize]{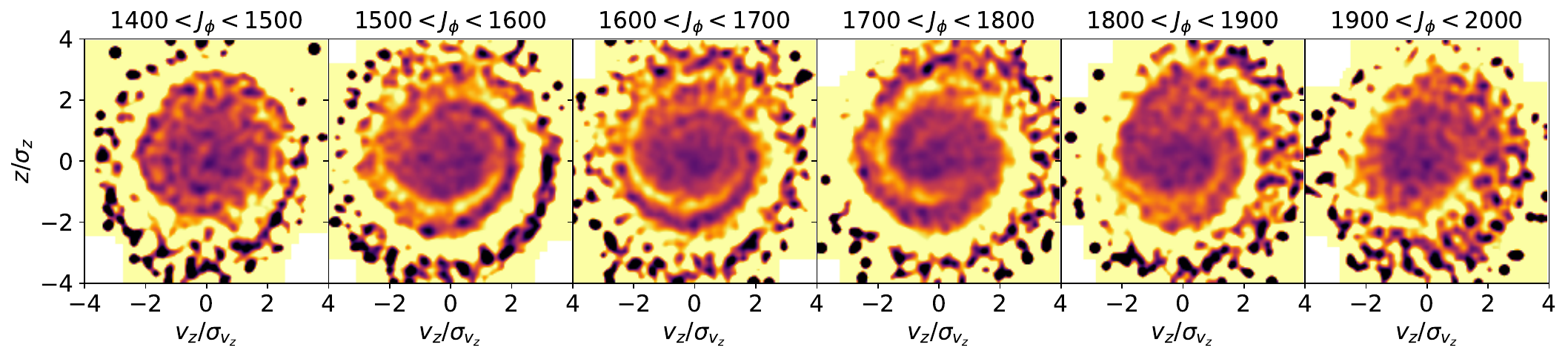}
    \includegraphics[width=\hsize]{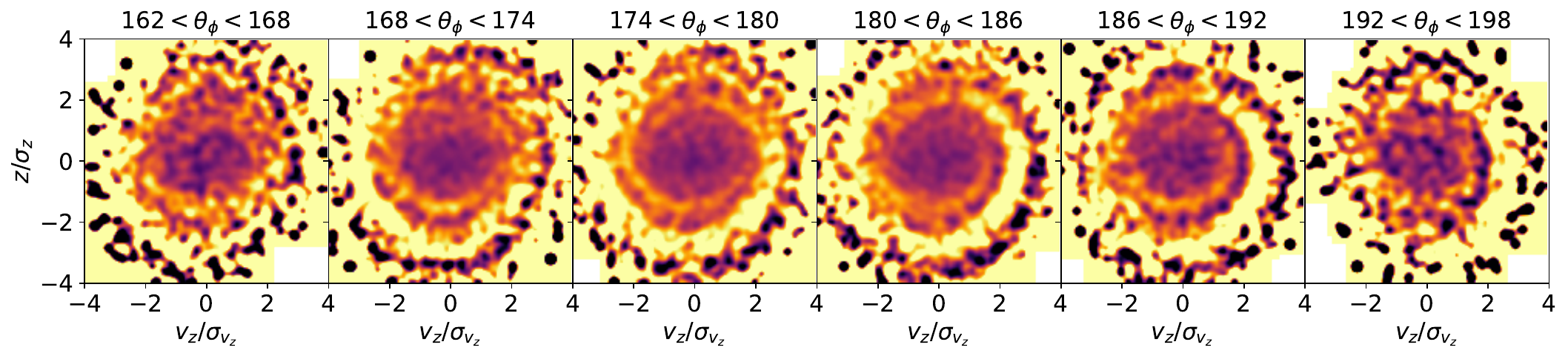}
    \caption{$z-v_z$ spiral for particles within 1 kpc of the `Solar location' at $(x,y,z)=(-8.178,0,0)$ kpc (upper left), within 3 kpc (upper middle) and within 5 kpc (upper right) in Model M1 at $t=6.874$ Gyr. The middle and lower rows are the upper left ($d<1$ kpc) selection of stars further subdivided into $J_{\phi}$ (middle row) and $\theta_{\phi}$ bins (lower row).}
    \label{M1-localspiral}
\end{figure}

As shown in \cite{Li20} and \cite{Gandhi+20}, the local phase snail as selected in spatial coordinates is a composite of phase snails with different angular momentum, $J_{\phi}$. The upper left panel of Figure \ref{M1-localspiral} shows the local phase spiral as defined as all stars within 1 kpc from the `Solar position' defined as $(x,y,z)=(-8.178,0,0)$ in model M1, where the $v_z$ and $z$ axes have been re-scaled by the vertical velocity dispersion $\sigma_{v_z}$ and the vertical dispersion $\sigma_z$ of the sample, respectively. For visualisation purposes the distribution is smoothed with a Gaussian with a scale of 0.2 in the re-scaled coordinates, and we have subtracted the smoothed background distribution, such that the colourbar shows the difference in density between the data and the smoothed version, $\Delta\rho$. The smoothing, scaling and color bar are kept consistent across all subsequent phase spiral plots. The upper middle and right hand panels show the $z-v_z$ spiral for selections within 3 and 5 kpc of the `Solar position' merely to illustrate the importance of being able to resolve the spiral locally. Selecting a larger region distorts the morphology, as expected.

\begin{figure*}
    \centering
    \includegraphics[width=\hsize]{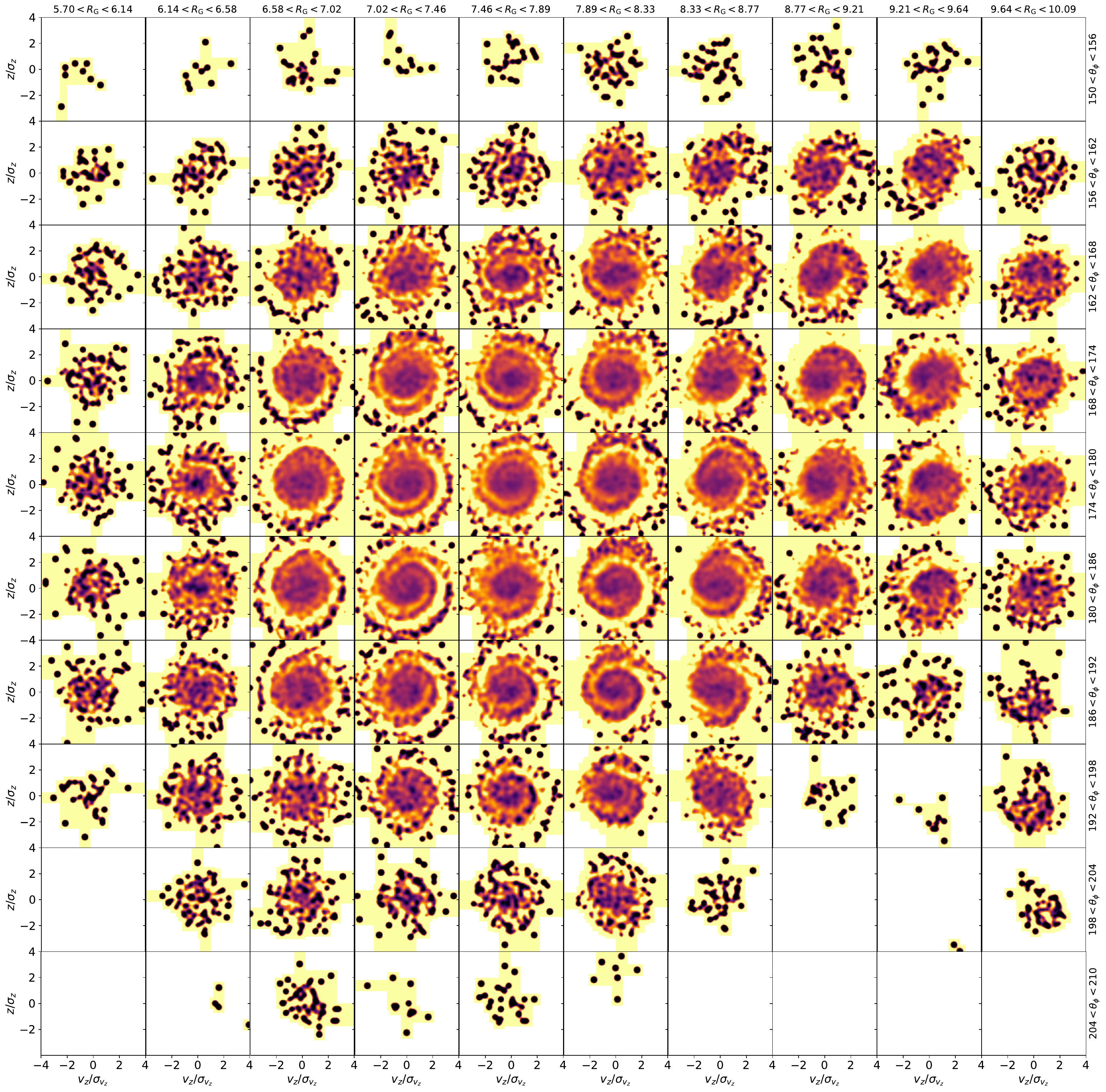}
    \caption{$z-v_z$ plane for a sample of stars within a 1 kpc box selection around the Solar neighborhood, split by $J_{\phi}$ (km s$^{-1}$ kpc, different columns) and $\theta_{\phi}$ (deg, different rows).}
    \label{bigsplit}
\end{figure*}

The middle row shows the split of stars from the $d<1$ kpc sample into different angular momentum bins. The morphology of the lower row evolves with $J_{\phi}$ as shown in \cite{Li20} and \cite{Gandhi+20}, because the phase spiral for stars selected purely in position space is an amalgamation of phase spirals from stars with different actions and different orbital histories which were located in different regions of the galaxy during the satellite perturbation \citep[see][for an illustration]{Gandhi+20}.

Furthermore, as shown in our previous work \cite{Hunt+20}, such a separation into different angular momentum bins, while retaining the physical azimuthal constraint is only half of the transformation from physical space into orbit space. Further refinement of the spiral should be possible by also selecting stars by their azimuthal phase angle, $\theta_{\phi}$, instead of their physical angle $\phi$, although this intrinsically assumes a form for the Galactic potential. $\theta_{\phi}$ is the angle of a star's guiding centre \citep[e.g. see Figure 1 of][]{Hunt+20} which increases linearly as it orbits the galaxy.

As such, the lower row of Figure \ref{M1-localspiral} shows the local phase spiral from the top panel split into $\theta_{\phi}$ bins (where $\theta_{\phi}=180$ deg is the Solar azimuth, and $\theta_{\phi}=0$ is the far side of the galaxy). As with the selection in $J_{\phi}$, the spiral evolves with $\theta_{\phi}$ for stars within the `local' volume. Similar to how stars with different actions are affected differently, so are stars with different azimuthal phase angles, because their initial position and motion with relation to the perturber will be different. This is to be expected, but has not been shown before to our knowledge.

Figure \ref{bigsplit} shows the phase spirals for the sample split by $J_{\phi}$ (displayed as $R_{\mathrm{G}}$) along the $x$-axis and by $\theta_{\phi}$ along the $y$-axis. Note that a 1 kpc volume around the sun covers $\sim7$ degrees in galactic azimuth, $\phi$, yet this expands to nearly 60 degrees in phase in the associated angle, $\theta_\phi$, while still resolving the phase spirals for around 30 degrees. Similarly, 1 kpc in galactocentric radius, $R$, contains stars for several kpc in $R_{\mathrm{G}}$ and lets us track the phase spiral for around 3 kpc radially. The morphology clearly evolves along both axes, and higher resolution models may be able to track the phase spiral even further from a local volume. Stars with the same $J_{\phi}$ but different $\theta_{\phi}$ show different morphology in the phase spiral, as do stars with the same $\theta_{\phi}$ but different $J_{\phi}$, as expected.

This splitting of phase spirals by stars with shared orbital histories further illustrates the variety of morphologies contained within the `Solar neighbourhood' phase spiral, and the significant amount of information which is lost when considering only the present day position of the stars and not their orbital histories. Analysing the phase spiral as a function of action-angle coordinates will enable us to probe the strength and timescale of a past impact at a range of positions around the galaxy, and may help us to constrain both the properties of the perturber, and the galactic potential, but we defer this to future work.

\subsection{Model M1: The phase spiral globally}\label{globalspiral}
A lot of attention has been paid to the local $z-v_z$ spiral since its discovery by \cite{Antoja+18}, but there also exists a spiral pattern in vertical position and motion which arises across the disc plane \citep[e.g.][]{Gomez+13,LMJG19,Gandhi+20}. It is difficult to resolve this global kinematic spiral in the actual Milky Way data, as we struggle to track the kinematics far from the Solar neighborhood \citep[although see][]{Eilers+20}, but there is no reason we can not examine it further in the simulations, and it is of course merely a different projection of the same vertical motion, and the same vertical asymmetry.

As a reminder, in Figure \ref{M1-global} the face on panels colored by mean $z$ (lower left) and mean $v_z$ (lower middle) show that the mean vertical position and motion of stars also exhibits spiral structure when viewed across the plane. These are a local collapsing of the phase spirals into their offset (or asymmetry) in either vertical position or velocity, albeit at a finer resolution than allowed when binning to resolve the phase spiral. However, with high resolution disc models we can resolve well defined phase spirals on the kpc scale all across the disc. 

As shown in Section \ref{localspiral}, it is useful to select the phase spirals as a function of $J_{\phi}$ and $\theta_{\phi}$. Thus, following \cite{Hunt+20}, we transform the physical coordinates $(x,y)$ into guiding centre Cartesian coordinates $(x_{\mathrm{act}},y_{\mathrm{act}})=(J_{\phi}/v_{\mathrm{circ}}\cos\theta_{\phi},J_{\phi}/v_{\mathrm{circ}}\sin\theta_{\phi})$, in order to group stars by their shared orbital history instead of their current physical location. Such a selection is not possible in the $Gaia$ data owing to selection effects, but it is straightforward in the simulation \citep[see][for a more thorough description of this transformation and its limitations]{Hunt+20}.

\begin{figure*}
    \centering
    \includegraphics[width=\hsize]{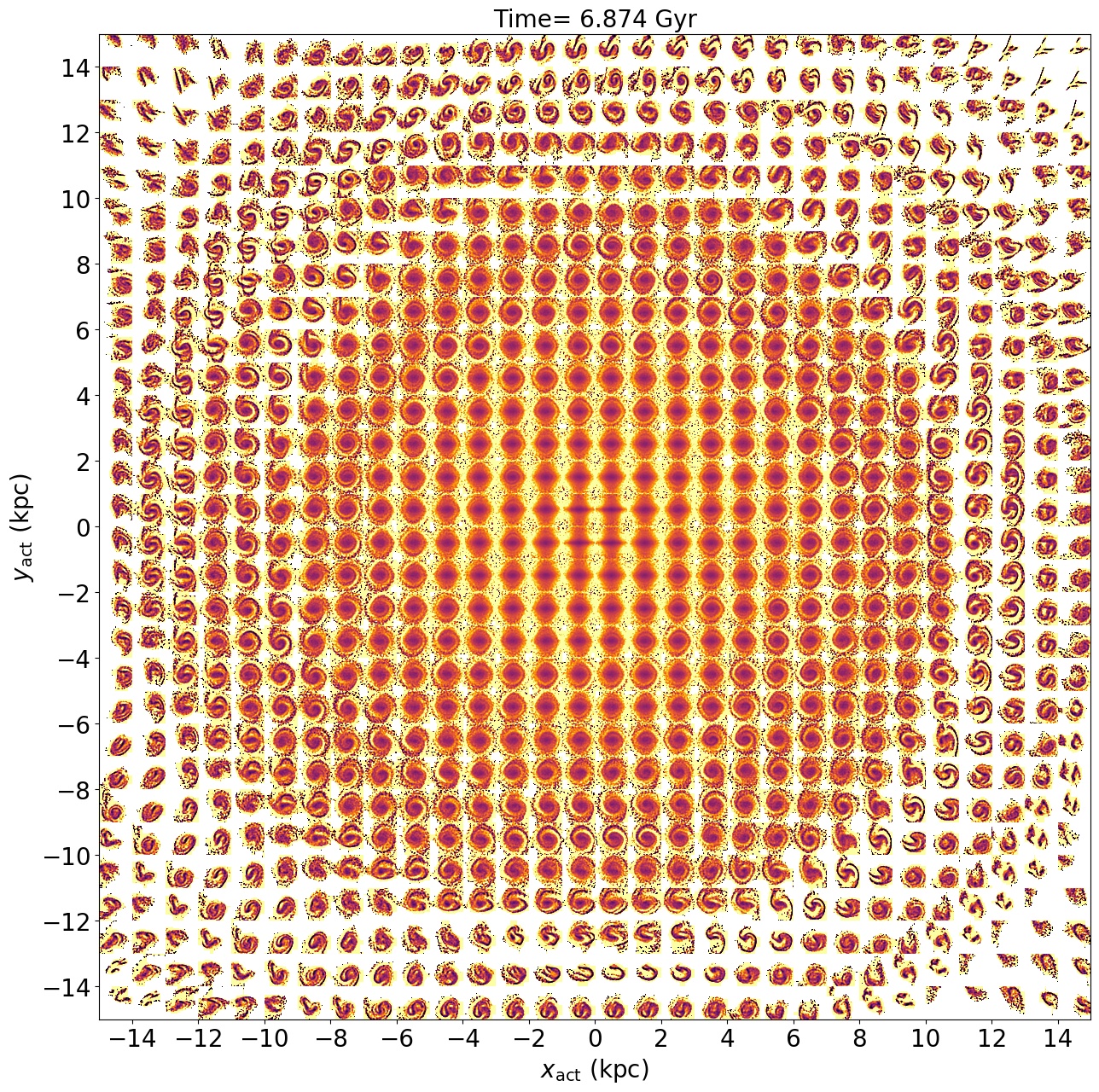}
    \caption{The phase spirals shown across the face of model M1 at $t=6.874$ Gyr, in $1\times1$ kpc bins from $X, Y=-15$ to 15 kpc. Note that the global vertical spiral is visible owing to the offset of the individual phase spirals from $(v_z,z)=(0,0)$ and the $v_z$ axes have been re-scaled to compensate for the change in velocity dispersion with radius (see text). For the animated version of the evolution over time see \url{https://users.flatironinstitute.org/~jhunt/M1-AASpiral.mp4} (or .gif if mp4 is unsupported).}
    \label{crazy1}
\end{figure*}

Figure \ref{crazy1} shows the face-on view of Model M1, in guiding centre Cartesian coordinates, divided into 1 kpc by 1 kpc bins, where each bin displays the phase spiral for the star particles contained within (see Figure \ref{M1-localspiral} for the axis scale and color bar for each individual spiral). Firstly, note that the $z-v_z$ phase spirals occur globally, not in an isolated region of the disc, but also note that their morphology and offset from the centre of each bin differs with location across the galaxy. This change in offset/asymmetry causes the larger spiral visible across the disc, as is shown in the vertical position and motion in Figure \ref{M1-global}. 

\begin{figure}
    \centering
    \includegraphics[width=\hsize]{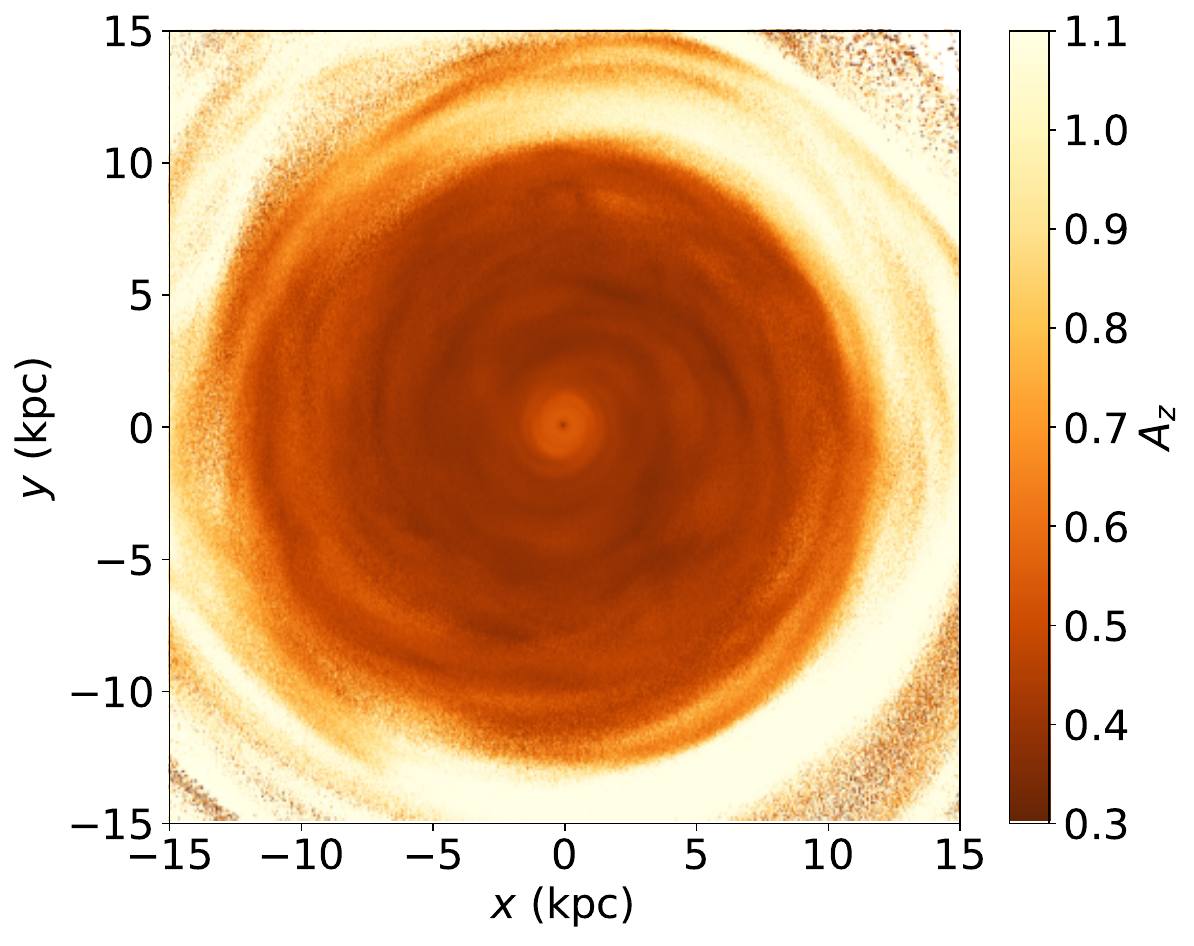}
    \caption{Face on map of vertical epicylcic amplitude, $A_z=(v_z/\Omega_z+z^2)^{1/2}$ for Model M1 present day snapshot, $t=6.874$ Gyr.}
    \label{azmap}
\end{figure}

Figure \ref{azmap} shows the face on view of the `present day' snapshot of Model M1, colored by $\bar{A}_z$, a simple model of the vertical epicyclic amplitude given by $A_z=(v_z/\Omega_z+z^2)^{1/2}$. The colourbar is such that the stars with the highest deviation from planar position or motion are in white, matching the offset of stars from the centre of the individual bins in Figure \ref{crazy1}, which traces out the global vertical spiral. The relation between the plane of phase spirals in the upper panel and the mean vertical epicyclic amplitude, which encompasses the offset in both axes is clear. 

However, Figure \ref{crazy1} is but a single snapshot in time, whereas a significant amount of information is contained within the time evolution of both the global and local vertical spirals. The time evolution of Figure \ref{crazy1} is available online\footnote{\url{https://users.flatironinstitute.org/~jhunt/M1-AASpiral.mp4} (or .gif if mp4 is unsupported)}. We encourage the reader to make use of the animation to visualise the propagation of the response to the repeated passages of the dwarf across the disc and the formation and evolution of the $v-v_z$ phase spirals. 

\begin{figure*}
    \centering
    \includegraphics[width=\hsize]{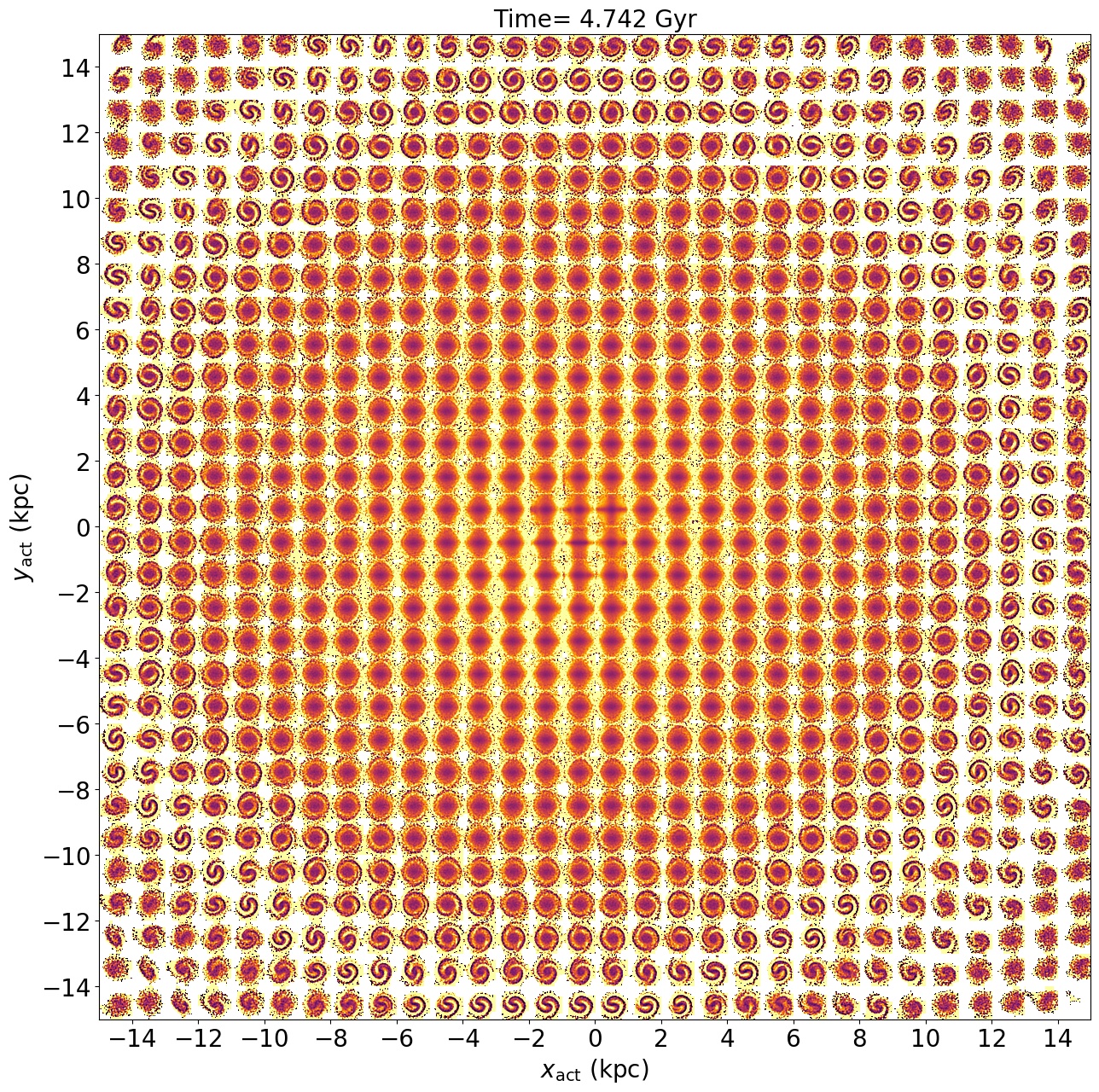}
    \caption{As Figure \ref{crazy1}, but for $t=4.73$ Gyr, the time of the second pericentric passage. The phase spirals currently present were excited by the first pericentric passage, 2.19 Gyr in the past.}
    \label{crazy2}
\end{figure*}

While the $z-v_z$ spirals close to the Solar circle are relatively well behaved in Figure \ref{crazy1}, those in the outer galaxy are more chaotic, being comprised of a mix of phase spirals induced by multiple passages \citep[see also][]{LMJG19}. The animated version of Figure \ref{crazy1}$^6$ shows that the first passage of the dwarf at $t=2.54$ Gyr causes only a small ripple across the global plane, but it is enough to perturb the disc from vertical equilibrium and seed the creation of the phase spirals, which propagate through the outer disc for over 2 Gyr before the second pericentric passage occurs at $t=4.74$ Gyr (see Figure \ref{orbits} for the orbit). Figure \ref{crazy2} shows as Figure \ref{crazy1}, but at $t=4.74$ Gyr, which is 2.2 Gyr after the first pericentric passage of the dwarf. The $z-v_z$ phase spirals around $R_{\mathrm{G}}=15$ kpc are well defined within the 1 kpc boxes following the single passage, highlighting the benefit of the increased phase space resolution of the disc.

\subsubsection{Bending and breathing behaviours, and the resulting phase-spirals}\label{bendbreath}
Interestingly, following the first passage, the phase spirals are two armed, as opposed to the single arms visible at later times (e.g. Figures \ref{M1-localspiral} to \ref{crazy1}, or the later stages of the animation). We argue that this is to be expected owing to the changing nature of the interaction between the disc stars and the dwarf over the course of the merger (see also \citealt{Poggio+2020,GrionFilho+20} for an exploration of the changing nature of the interaction between the dwarf and satellite on a global scale in the L2 model of \citealt{LJGG-CB18}).

The first passage occurs at a large radius ($\sim35$ kpc), with high vertical velocity ($\sim350$ km s$^{-1}$), which is a rapid, almost entirely perpendicular passage through the outer part of the galaxy, where the satellite velocity is much greater than the maximum vertical velocity of the stars ($\sim80$ km s$^{-1}$ at $\geq15$ kpc). The interaction timescale is on the order of $h_{\mathrm{d}}/v_{z,\mathrm{sat}}= 0.439\ \mathrm{kpc}/350\ \mathrm{km\ s}^{-1}\approx1.2$ Myr, which is significantly faster than the vertical period of the orbits (see Figure \ref{freqs}). 

The rapid transit also leaves little time for the stars to travel azimuthally, meaning that overall they receive an approximately equal force upwards and downwards. As previously shown in \cite{Widrow+14}, such an interaction excites a breathing motion in the disc, owing to the short timescale of the interaction, and the high relative velocity between the perturber and the disc stars. 

Conversely, the later passages have increasingly longer interaction times and a smaller difference in vertical velocity. For example, the second and third disc crossings are close together in time ($t=4.40$ and $t=4.85$, with the pericentre in between). The interaction timescales here are on the order of $R_{\mathrm{cross}}/v_{z,{\mathrm{sat}}}=74\ \mathrm{kpc}/130\ \mathrm{km\ s}^{-1}\approx557$ Myr and $R_{\mathrm{cross}}/v_{z,{\mathrm{sat}}}=38\ \mathrm{kpc}/230\ \mathrm{km\ s}^{-1}\approx162$ Myr for the second and third disc crossings respectively, and the orbital path is such that the dwarf exerts a continuous influence on the disc while passing below the galaxy and remaining comparatively close to the disc plane for an extended period of time (see Figure \ref{orbits}). Such interactions on the hundreds of Myr scale are then comparable to the disc frequencies (see Figure \ref{freqs}), and should induce bending waves \citep[which would be strongest when the satellite is in resonance with the disc stars;][]{SNT98}.

The behaviour of the stars themselves is also dependent on the vertical phase angle of their orbit, $\theta_z$. In brief, in a high velocity, rapid interaction, stars with $\theta_z=0$ or $\pi$ (apocentres) both gain energy, while stars with $\theta_z=\pm\pi/2$ (pericentres) will lose energy, stretching the $z-v_z$ phase space distribution and creating a breathing motion. Conversely, in a slower interaction, for two stars spaced by $\pi$ in angle, one will gain energy and one will lose energy, resulting in an overall shift in the distribution and the creation of a bending motion \citep[see Section 2 of][for a thorough explanation and illustration]{Widrow+14}.

To our knowledge this has not been shown in other MW-Sgr merger studies, which have typically focused on the latter stages of the merger \citep[although see][]{LJGG-CB18}, by which point the interaction is on the timescale and relative velocity to create bending modes \citep[e.g.][]{BH-TG+21} and thus the one armed spirals. Thus, it is possible that relics of such two armed `breathing spirals' could exist in the outer disc of the Milky Way (e.g. as seen at around $R_{\mathrm{G}}=13$ kpc in Figure \ref{crazy1}), and they could be resolvable with future surveys. While the orbital trajectory of Sgr is not well known over timescales of several Gyr it is likely that early interactions were relatively quick, and the disc should contain both bending and breathing motions \citep[e.g.][]{Widrow+14}. As shown in \cite{LMJG19}, while subsequent passages 
of a satellite can erase the signatures of previous perturbations in the Solar neighbourhood, the longer orbital timescales in the outer disc make it more likely that multiple patterns can coexist. Regardless, we consider this an example of an interesting dynamical phenomenon brought to light by the high phase space resolution in the outer disc.

\subsection{Model M1: Planar dynamics and the classical moving groups}\label{UV}
\begin{figure}
    \centering
    \includegraphics[width=\hsize]{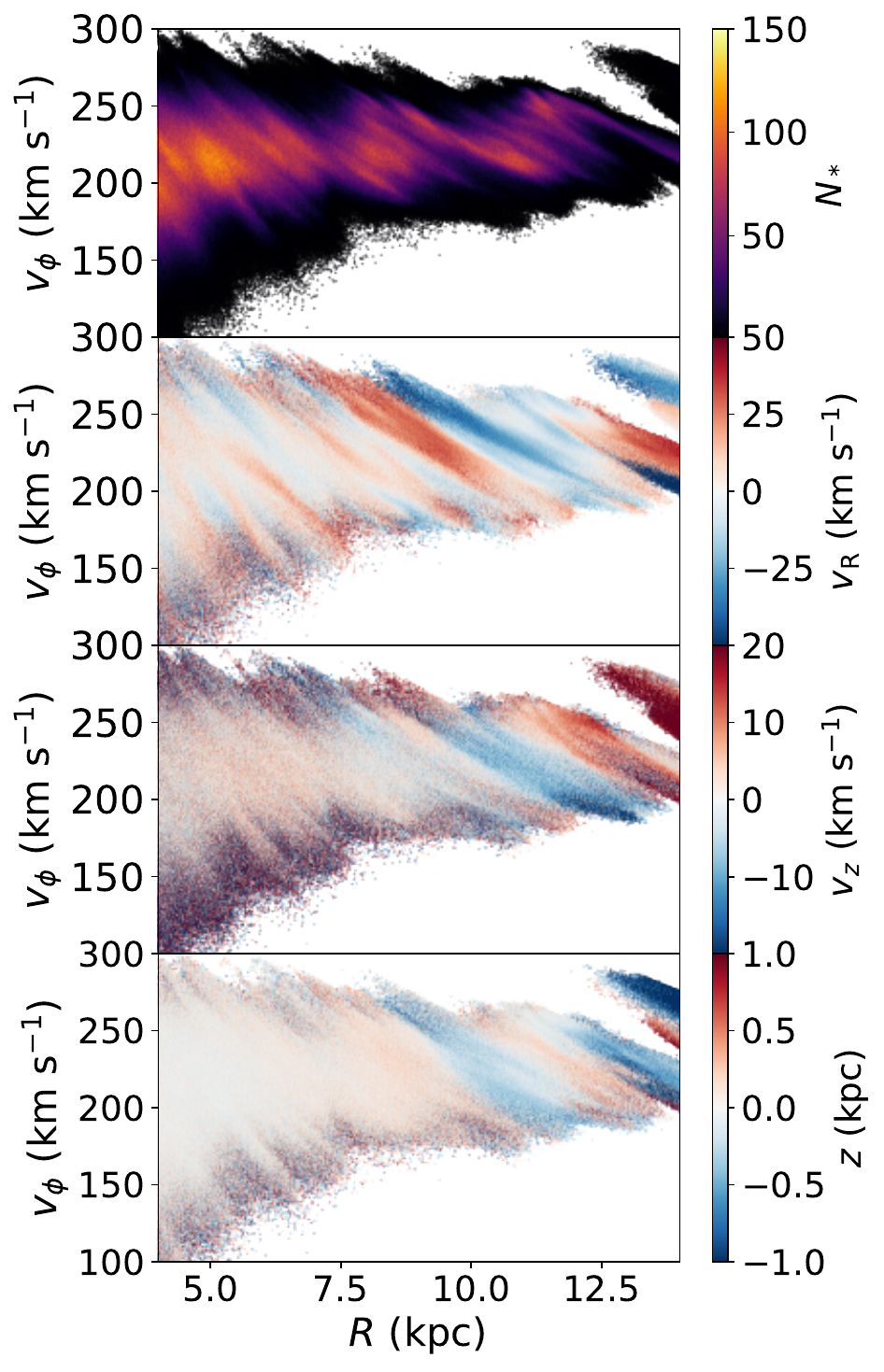}
        \caption{$R-v_{\phi}$ distribution colored by number count (top panel), mean radial velocity (km s$^{-1}$, second panel), mean vertical velocity (km s$^{-1}$, third panel) and mean vertical position (kpc, bottom panel) for model M1 at the `present day' snapshot.}
    \label{M1-ridges}
\end{figure}

While most of the recent interest in the impact of Sgr upon the disc of the Milky Way has been regarding the vertical kinematics, it has also been shown that such mergers reproduce similar ridge structure as are present in the $R-v_{\phi}$ plane \citep{LMJG19} and arches in the $v_{\mathrm{R}}-v_{\phi}$ plane \citep{Khanna+19}which were previously predicted in \cite{minchev09}. Figure \ref{M1-ridges} shows the $R-v_{\phi}$ plane for the `present day' snapshot of Model M1 for star particles with $-\pi/32<\phi<\pi/32$, colored by number density (top panel), radial velocity (second panel), vertical velocity (third panel) and vertical position (bottom panel). As with previous studies, it is unsurprising to find significant structure in this space, despite the lack of a galactic bar, which is often invoked to create such features through resonant interaction with the disc \citep[e.g.][]{Fragkoudi+19,Asano+20}. The magnitude of the radial velocity and vertical velocity pattern are both approximately double that observed in the $Gaia$ data, but this is to be expected owing to the high mass of the dwarf remnant in Model M1.

The real advantage of having high phase space resolution in the disc is that we can resolve the $v_{\mathrm{R}}-v_{\phi}$ plane over a smaller local volume than previous studies \citep[e.g.][]{Khanna+19}. While we still cannot match the resolution of the $Gaia$ data, we can easily pick out several moving groups when selecting regions on the kpc or less scale at the Solar radius. For example, \cite{FBBPZ18} have previously used \texttt{Bonsai} simulations to examine the $v_{\mathrm{R}}-v_{\phi}$ plane around the Outer Lindblad Resonance (OLR) of a short fast bar. Similar to previous studies \citep[e.g.][]{D00} they find it causes a Hercules-like stream, a well known moving group in the Solar neighbourhood, but only in approximately 50\% of snapshots. At other times it is disrupted or concealed by interaction with galactic spiral structure. 

\begin{figure}
    \centering
    \includegraphics[width=\hsize]{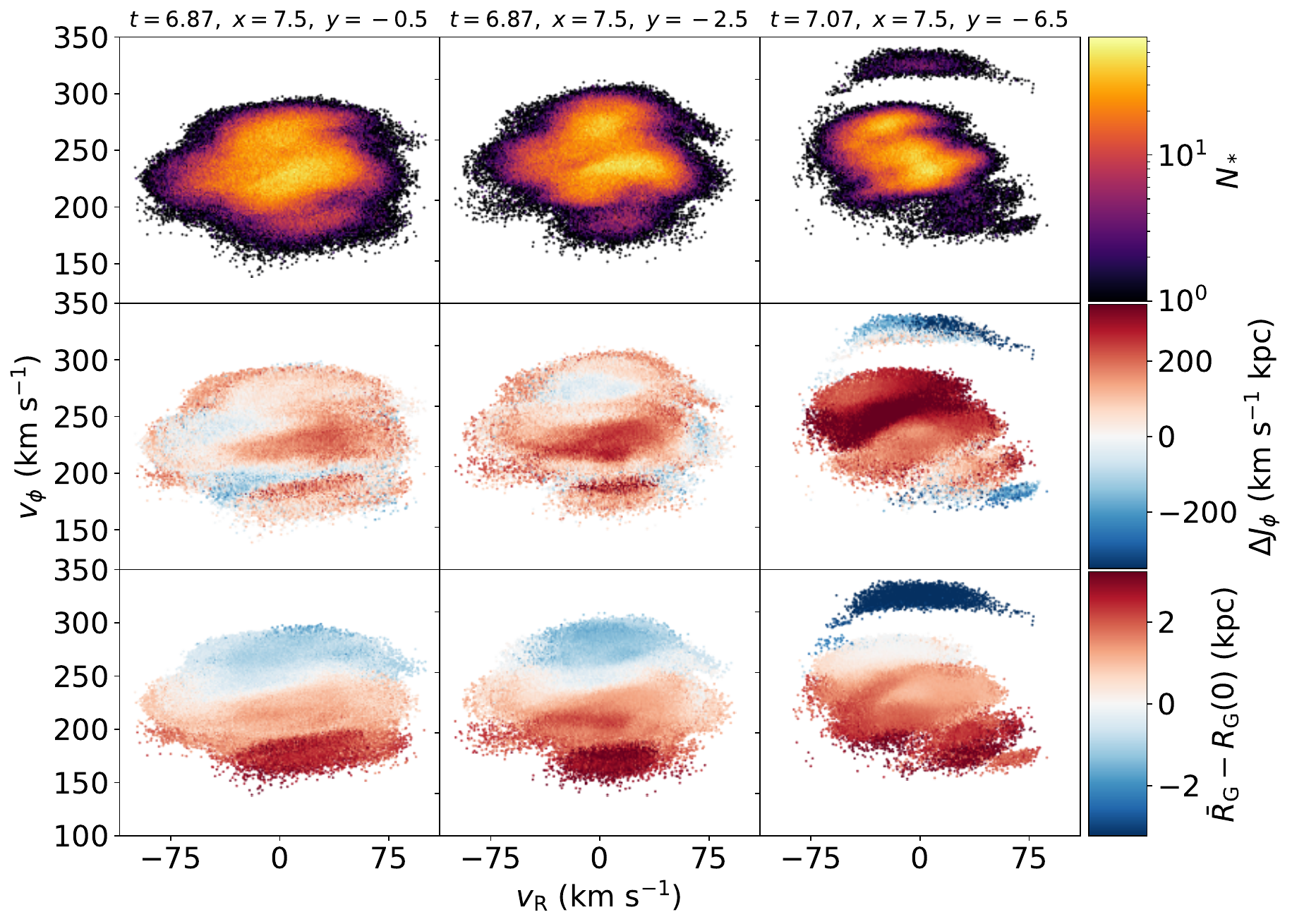}
    \caption{$v_{\mathrm{R}}-v_{\phi}$ distribution colored by number count (upper row), change in angular momentum, $\Delta J_{\phi}$ (km s$^{-1}$ kpc) from the initial condition (middle row) and initial guiding radius with respect to the current location of the sample (lower row) for three local samples in the `present day' snapshot (left and middle column) and 200 Myr later (right column).}
    \label{uvs}
\end{figure}

However, as mentioned above, numerous recent measurements of the bar pattern speeds favour a slower bar of around 40 km s$^{-1}$ kpc$^{-1}$ \citep[e.g.][]{Portail+17,BLHM+19,SSE19,Clarke+19}, which would put the OLR a few kpc outside the Solar neighbourhood, and no longer possible to cause the Hercules stream. Instead, a `slow' bar with pattern speed around 40 km s$^{-1}$ kpc$^{-1}$ will influence the local kinematics in the region of Hercules, through either the corotation resonance \citep{P-VPWG17}, or the 4:1 OLR \citep{HB18}. However, the corotation resonance of a reasonable strength bar generally produces a significantly weaker Hercules-like stream than is seen in the data \citep[e.g.][]{Binney2018,Hunt+18}, although see also \cite{Monari+19} who use the \cite{Portail+17} potential. Similarly, if Hercules is caused by the 4:1 OLR, then we would expect a strong OLR feature in a region of velocity space which is not observed in the Solar neighbourhood \citep[for a thorough discussion of the kinematic signature of bars with fixed pattern speeds see][and references within]{Hunt+19,Trick+19}. 

Alternatively, \cite{CFS19} proposed the `resonance sweeping' of a slowing bar as a mechanism for the creation of the Hercules stream and other Solar neighbourhood kinematics, an appealing explanation which can reconcile direct measurements of the Galactic bar's slow pattern speed with the strength of the kinematic substructure induced by the corotation resonance observed in the Solar neighbourhood. More recently, \cite{CS21} show that such a slowing bar creates a Hercules-like stream which is metal-rich by 0.1 to 0.2 dex, owing to the constituent stars originating in the inner galaxy. 

We do not intend or expect to match the Solar neighborhood $v_{\mathrm{R}}-v_{\phi}$ plane in Model M1, which is a model with no bar, and a disc with $Q=2.3$. However, similar to previous studies of the impact of spiral structure on the local velocity distribution \citep{Hunt+18,Hunt+19}, we find it straightforward to locate regions with a Hercules like feature in the appropriate part of velocity space in the model prior to bar formation. It is not straightforward to examine the metallicity of this region as model M1 is a pure $N$-body simulation without gas or star formation. However, we can track the radial migration the star particles experience during the merger to see if our Hercules-like streams originate in the inner galaxy, which we would expect to be more metal rich.

The top row of Figure \ref{uvs} shows the $v_{\mathrm{R}}-v_{\phi}$ planes for three regions close to the Solar radius in the present day snapshot of Model M1 (left two columns) and one from 175 Myr afterward which bares a striking resemblance to the actual Solar neighborhood $v_{\mathrm{R}}-v_{\phi}$ plane (right column). Each has a small separate moving group with low $v_{\phi}$ and positive $v_{\mathrm{R}}$, qualitatively similar to the Hercules stream in the Solar neighborhood. The second row shows the $v_{\mathrm{R}}-v_{\phi}$ plane colored as a function of the mean change in angular momentum, $\Delta J_{\phi}$ (km s$^{-1}$ kpc), since the start of the simulation. The third row shows the difference between the initial guiding radius, $R_{\mathrm{G}}(0)=J_{\phi}(0)/v_{\mathrm{circ}}$, which is a proxy for their birth radius, which is in turn a proxy for their metallicity, relative to the current guiding radius of the sample of the star particles. 

\begin{figure*}
    \centering
    \includegraphics[width=\hsize]{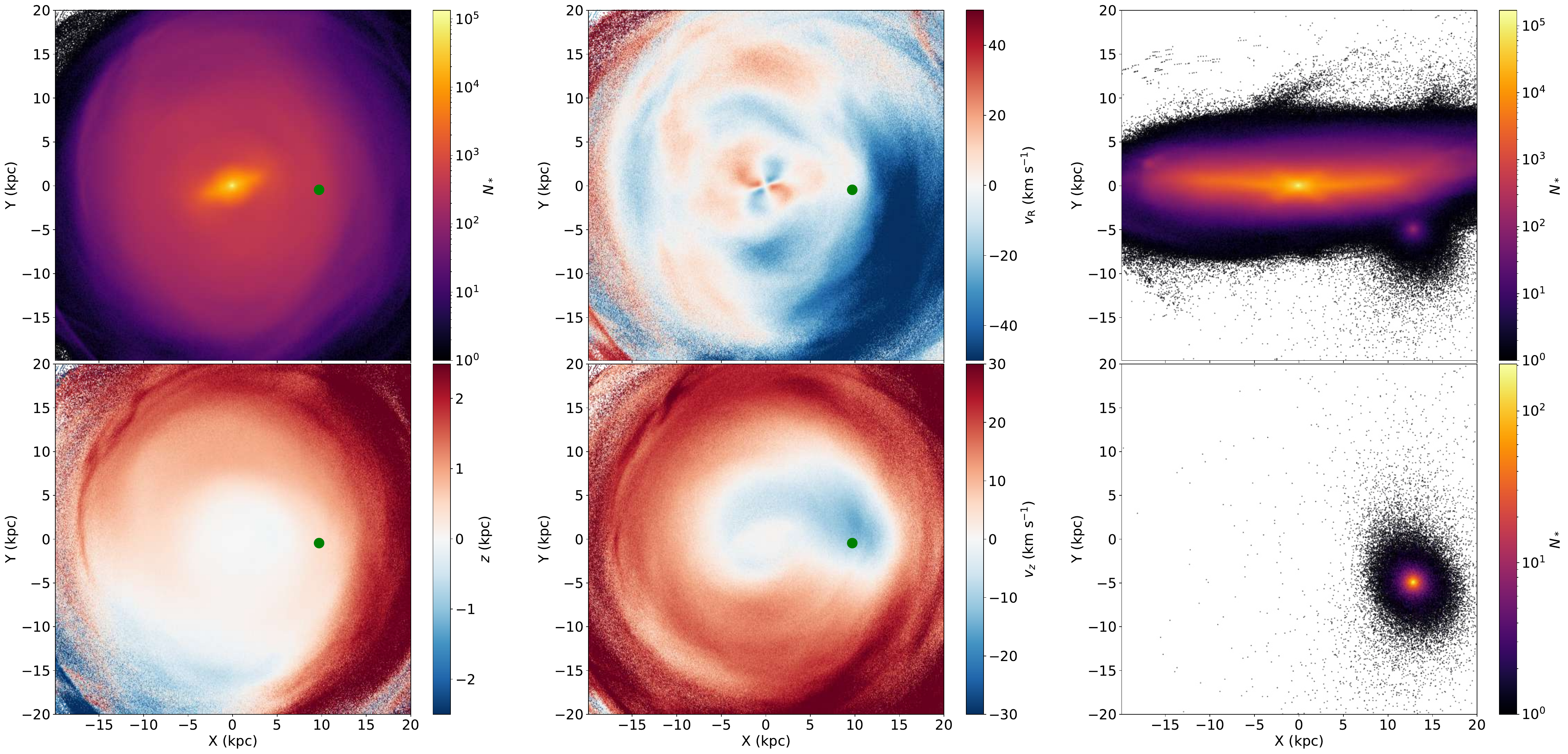}
    \caption{Face on (left and centre columns) and edge on (right column) view of the disc colored by number count (upper left panel and right column), mean radial velocity (upper centre), mean vertical position (lower left) and mean vertical velocity (lower centre) for Model M2 at $t=5.44$ Gyr. The dwarf particles are only included in the right hand column, but its position is marked with a green dot in the face on panels.}
    \label{M2-global}
\end{figure*}

\begin{figure*}
    \centering
    \includegraphics[width=\hsize]{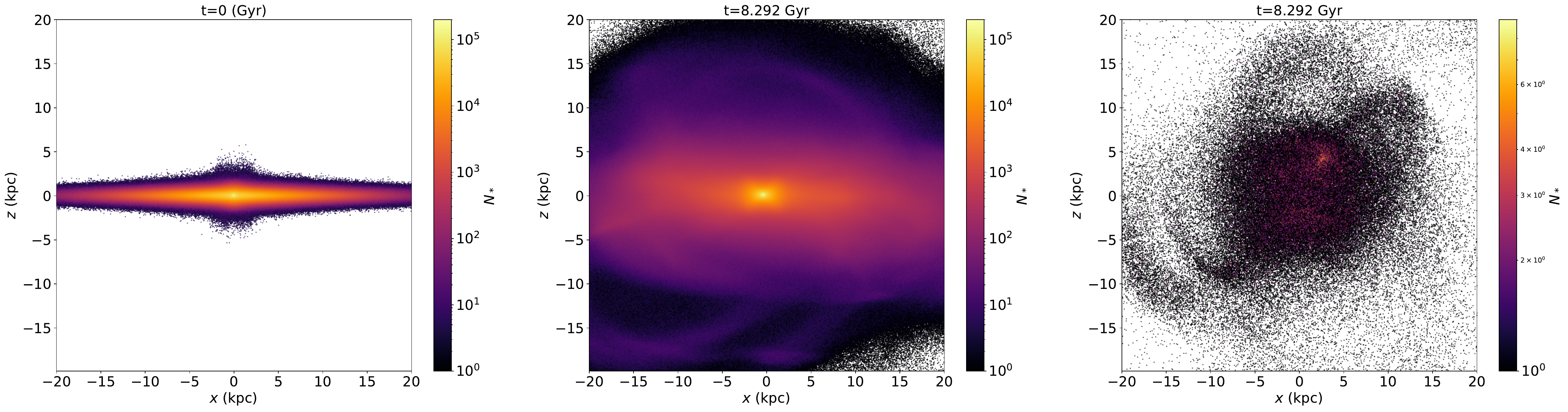}
    \caption{Edge on view of Model M2 at $t=0$ (left panel) and $t=8.292$ (Gyr, middle panel), and just the accreted stars from the dwarf at $t=8.292$ (right panel)}
    \label{M2-finalZ}
\end{figure*}

\begin{figure}
    \centering
    \includegraphics[width=\hsize]{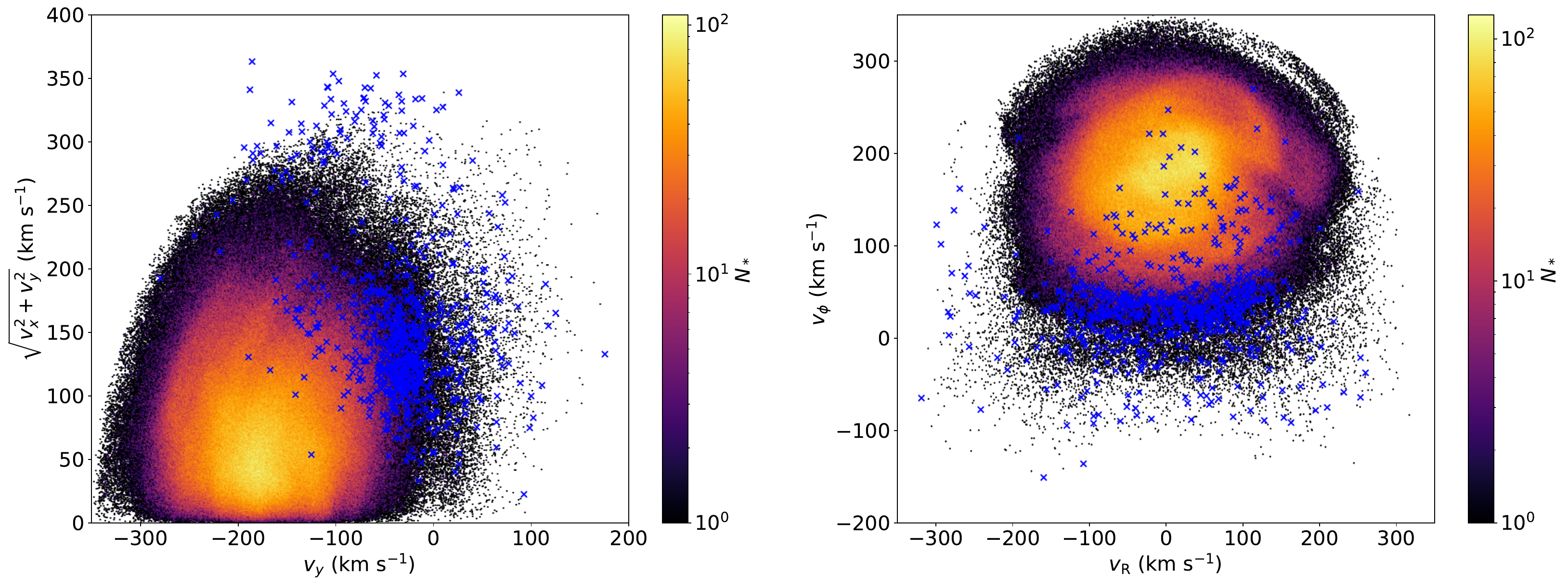}
    \caption{Toomre diagram (left) and the $v_{\mathrm{R}}-v_{\phi}$ plane (right) for the disc stars (histogram) overlayed with the accreted stars from the dwarf (blue crosses) both selected within 2 kpc of a Solar-like position of $(x, y, z)=(-8.178,0,0)$ kpc.}
    \label{sausage}
\end{figure}

In each case we see that the Hercules-like moving group is mainly comprised of `red' stars from the inner galaxy which have migrated outwards, with low initial $R_{\mathrm{G}}(0)$, and positive migration, $\Delta J_{\phi}>0$. Hence, the Hercules-like group in each case would be more metal rich. In addition, the largest implied metallicity occurs in the centre of the outwards moving group, similar to as found by \cite{CS21} for the slowing bar model, while we also see that the `horn' feature present in the right hand column is `metal rich', as seen in the data, but not in the slowing bar model.

We are certainly not claiming that the Milky Way does not contain a bar, or that it does not impact the kinematics of the disc stars. Both those statements are undoubtedly false, but as discussed before \citep{Hunt+19} we remind the reader that it remains dangerous to make measurements of bar length, strength or pattern speed from a feature which can be explained by a different origin, or at the very least could be distorted by other non-axisymmetric structure. 

In the case of the real data, the local $v_{\mathrm{R}}-v_{\phi}$ distribution is likely to have been shaped by a combination of the Galactic bar, spiral structure and the perturbation of the disc caused by the interaction of the Milky Way and its satellite galaxies, and disentangling them based on phase-space information and metallicity alone is non-trivial. High resolution simulations containing a Milky Way-like galactic bar, spiral structure and satellite perturbers may help us break the degeneracy.

\section{Model M2 structure and kinematics}\label{M2}
Figure \ref{M2-global} shows the face and edge on morphology and kinematics for Model M2, as detailed above for Figure \ref{M1-global}. In this snapshot, the dwarf is a little closer to the true location of Sgr in the Milky Way than Model M1. However, it is significantly less disrupted and has barely started to form a tidal stream making it a worse model of the overall system. In addition, as the dwarf remains significantly more concentrated, the force acting on the disc by the more massive remnant causes significant perturbation even at this early stage of the Merger, creating a strong warp. By the time the dwarf starts to disrupt it is further inside the Solar circle (see Figure \ref{orbits}), and has caused significant perturbation to the disc beyond what is observed in the Milky Way. For this reason we do not consider the `present day' snapshot of Model M2 defined by the position of the dwarf to be appropriate for comparison with our own Galaxy in respect to the current interaction of the Milky Way and Sgr. Instead, it is interesting as a high-resolution model of a more violent merger.

For example, Model M2 is potentially similar to violent merger events much earlier in the Milky Way's accretion history. Such mergers have been proposed to be responsible for the creation of the Milky Way's thick disc \citep[e.g.][]{QHF93}, although other explanations exist, such as the merger of numerous small satellites \citep[e.g.][]{ANSE03}, in-situ formation following a gas rich merger \citep[e.g.][]{BKGF04}, the heating of the disc by radial migration \citep[e.g.][]{SB09} or inheritance at birth through the properties of turbulent gas at high redshift \citep[e.g.][]{noguchi98, Bournaud+09}. Figure \ref{M2-finalZ} shows the edge on view of Model M2 at $t=0$ (left panel), and $t=8.29$ (Gyr, middle panel), and the accreted dwarf stars at $t=8.29$ (right panel)\footnote{Model M2, snapshot 557 on Flathub}. The disc clearly experiences a significant thickening during the merger, and there remains numerous structures in the heated disc. The dwarf stars are well dispersed within $\sim5$ kpc, with some stream-like structure remaining in the outer galaxy. The lack of gas in Model M2 will prevent any subsequent formation of a replacement thin disc, but we can examine the thickened stellar disc and newly formed inner halo, and probe the dynamics of the accreted stars.

Figure \ref{sausage} shows the Toomre diagram \citep[left;][]{SF87} and the $v_{\mathrm{R}}-v_{\phi}$ plane (right) for the disc stars (histogram) overlayed with the accreted stars from the dwarf (blue crosses) both selected within 2 kpc of a Solar-like position of $(x, y, z)=(-8.178,0,0)$ kpc. The left panel shows the local accreted stars (blue crosses) on more radial orbits slightly outside the main disc volume (histogram). This is similar to Figure 1 of \cite{Helmi+18} comparing their merger simulation to the Gaia-Enceladus stars. Note that our merger is slightly prograde instead of slightly retrograde, as the model was not tailored to the Gaia-Enceladus merger event. Similarly, the right panel shows the accreted stars on radial orbits reproduce the `sausage' like structure as shown in \cite{BEEKD18}.

An $N$-body model of an isolated system is of course a less effective representation of the long term evolution of a Milky Way-like galaxy when compared to cosmological simulations that contain a full merger history while also encapsulating additional physics such as star formation and feedback. However, such simulations are, by necessity, lower resolution than isolated discs \citep[although see][]{Grand+21} and we make the final step of model M2 available as an example of a high resolution violent merger, which despite the lack of gas, nonetheless captures the satellite disruption and impact on the pre-existing disc.

\section{Conclusions}\label{conclusions}
In this work, we present two $>1$ billion particle self consistent simulations of a dwarf galaxy merging into a Milky Way-like disc galaxy, and the two isolated galaxies for comparison. 

Model M1 is a merger into a highly stable disc (D1, $Q=2.3$) where the non-axisymmetric structure all arises from the merger, making it a good laboratory to investigate the response. Model M2 is a more violent merger into a more massive disc (D2) which experiences significant disruption similar to mergers earlier in the Milky Way's formation history, which produces a thick disc and an accreted population qualitatively similar to stars from the Gaia-Enceladus merger, both in the Toomre diagram and the `Sausage' in the $v_{\mathrm{R}}-v_{\phi}$ plane. All are freely available on Flathub\footnote{\url{https://flathub.flatironinstitute.org/jhunt2021}}.

Using Model M1 we illustrate the ability to resolve both vertical and planar kinematic substructure over a local volume, and show the emergence and evolution of the $z-v_z$ phase spirals globally across the disc. Our three main conclusions are:

i) We observe for the first time the different morphology of the $z-v_z$ phase spirals which arise from a breathing motion excited by an early rapid disc crossing (two armed spiral) compared to those which arise from bending modes (one armed spiral) during later slower disc crossings.

ii) We show that these relics of earlier disc crossings are seen for multiple Gyr, especially in the outer Galaxy, and shorter simulations of only the latter stages of such a merger may miss some complexity in both the resulting $z-v_z$ phase spirals and the global dynamics. Such relics could act as fossils into prior perturbations and/or laboratories to studying the growth rate of the Galaxy \citep[e.g.][]{laporte20}. 

iii) We show that Model M1 creates numerous Hercules-like moving groups in local planar kinematics, despite the lack of tailoring of the simulation to the Milky Way or the presence of a Galactic bar. These moving groups originate in the inner galaxy, and would be metal rich compared to the rest of the $v_{\mathrm{R}}-v_{\phi}$ plane, as observed in the Solar neighbourhood data. 

Finally, we remind the reader that while these models are an interesting high resolution exploration of merger dynamics, they are not tailored to the MW-Sgr system, or the Gaia-Enceladus merger. For future work we intend to create and release simulations of more Milky Way-like systems, merging with more Sgr-like dwarf impactors \citep{BBH21}, and also include the Magellanic clouds, which are likely important for the present day state of the overall system \citep[e.g.][]{Gomez15,Laporte+18a,Nico+19,Cunningham+20,Vasiliev+21}.

\section*{Acknowledgements} 
We thank the referee for a constructive report. JASH is supported by a Flatiron Research Fellowship at the Flatiron institute, which is supported by the Simons Foundation. KVJ was supported by NSF grant AST-1715582. CFPL acknowledges funding from the European Research Council (ERC) under the European Union's Horizon 2020 research and innovation programme (grant agreement No. 852839). This work was supported in part by World Premier International Research Center Initiative (WPI Initiative), MEXT, Japan. This research made use of \texttt{astropy}, a community-developed core Python package for Astronomy \citep{astropy-1,astropy-2}, and the galactic dynamics Python packages \texttt{galpy} \citep{B15}, \texttt{Agama} \citep{agama} and \texttt{Gala} \citep{gala}. This work was performed in part at Aspen Center for Physics, which is supported by National Science Foundation grant PHY-1607611. This work was partially supported by a grant from the Simons Foundation. 

\section*{Data availability}
The simulations are available on Flathub, at \url{https://flathub.flatironinstitute.org/jhunt2021}.

\bibliographystyle{mn2e}
\bibliography{ref2}

\label{lastpage}
\end{document}